\documentclass[%
 reprint, 
 amsmath, amssymb, 
 aps, 
]{revtex4-2}

\usepackage{graphicx}
\usepackage{dcolumn}
\usepackage{bm}

\begin{document}

\preprint{APS/123-QED}

\title{Skyrmion Generation through the Chirality Interplay of Light and Magnetism}

\author{Qifan Zhang\(^{1,2,3}\)}%
\author{Shirong Lin\(^{2,3,}\)}
\email{shironglin@gbu.edu.cn}
\author{Wu Zhang\(^{1,}\)}
\email{zhangwu@gzhu.edu.cn}

\affiliation{\makebox[0pt][l]{\textsuperscript{1}} School of Physics and Materials Science, Guangzhou University, Guangzhou 510006, China}%

\affiliation{\makebox[0pt][l]{\textsuperscript{2}} School of Physical Sciences, Great Bay University, Dongguan 523000, China}

\affiliation{\makebox[0pt][l]{\textsuperscript{3}}  Great Bay Institute for Advanced Study, Dongguan 523000, China}




\date{\today}

\begin{abstract}
Light beams, with their rich degrees of freedom, including polarization and phase, along with their flexible tunability, have emerged as an ideal tool for generating magnetic topological textures. However, how to precisely control the light beams to generate a specific number of magnetic topological textures on demand remains a critical scientific issue that needs to be resolved. Based on the numerical simulation of the Landau-Lifshitz-Gilbert equation, we propose that circularly polarized Laguerre-Gaussian beams can induce chiral magnetic fields through the interaction of the chirality of these beams' angular momenta. By utilizing these chiral magnetic fields, skyrmions or skyrmionium can be induced in chiral magnets. Moreover, the vectorial magnetic fields can be manipulated by adjusting the angular momenta and light intensity, thereby generating target chiral patterns and strengths, which allows for precise control over the type and number of these topological magnetic textures. This finding not only reveals the underlying physical mechanisms of the interaction between light and magnetic systems but also provides a feasible solution for the on-demand generation and encoding of skyrmions. 
\\

\end{abstract}

\maketitle


\section{Introduction}

Recent advances in light-matter interactions, involving structured light \cite{au2013direct,blanco2018ultraintense,fujita2019accessing,lin2022all} or engineered materials \cite{fu2020optical,fu2020optical,graf2021design,lin2023controllable,liang2024local}, have led to the realization of a wealth of novel physical phenomena, promoting the interdisciplinary applications of optics and other fields. In particular, light has degrees of freedom, including polarization and phase, which can manifest as two types of angular momentum, and its carrying of angular momentum endows it with unique properties, adding many possibilities to the interaction between light and matter. In 1936, Beth et al. experimentally demonstrated that circularly polarized light has spin angular momentum (SAM), which is generated by the spin of photons \cite{ beth1936mechanical}. Later, in 1992, Allen et al. experimentally showed that Laguerre-Gaussian (LG) light beams carrying orbital angular momentum (OAM) possess a helical phase factor of the form \(\mathrm{exp}(im\varphi)\), where \textit{m} represents the topological charge, and \(\varphi\) is the azimuthal angle of the light beams \cite{allen1992orbital}. Light beams carrying OAM are typically referred to as vortex beams. Vortex beams have phase singularities, with the phase distributed in a helical pattern along the radial direction. Since the late 20th century, research on vortex beams has developed rapidly, and their unique properties have shown great potential for applications in optical manipulation, quantum information and communication, optical imaging, and optical sensing. A considerable amount of research has been dedicated to exploring the functions of SAM and OAM with a wide range of applications, including in the field of magnetism \cite{lin2019all, nakata2019laser, banerjee2022inverse,guan2023optically, ghosh2023ultrafast, ono2023photocontrol, joseph2024role, mochizuki2012spin, finazzi2013laser, osada2018brillouin, fanciulli2021electromagnetic, goto2021twisted, watzel2022light, fanciulli2022observation, gao2023dynamical, rameshti2024twisted,du2019deep,sanchez2023all}. 

Skyrmions are topological solitons that can also exist in magnetism \cite{roessler2006spontaneous, mühlbauer2009skyrmion}, and are characterized by a swirl-like spin texture. Their stable topological configurations make them highly promising for applications in magnetic storage and magnetic encoding technologies \cite{nagaosa2013topological}. Skyrmions can be generated in various ways, including via electric current \cite{mochizuki2015writing,yuan2016skyrmion}, spin waves \cite{liu2015skyrmion}, acoustic waves \cite{yokouchi2020creation} and light \cite{ogawa2015ultrafast, buttner2021observation,koshibae2014creation}. 
The rapid writing and erasing of skyrmions, as well as their stable manipulation, are key factors for the application of skyrmions in magnetic storage \cite{guan2023optically}. Conventional approaches for skyrmion control encompass current-driven \cite{liu2013mechanism}, anisotropy gradient-driven \cite{guan2022unidirectional}, spin wave-based \cite{guan2021suppression, li2018dynamics, zhang2015all} as well as thermally activated manipulation \cite{kong2013dynamics,  kong2021dynamics, qin2022dynamics, raimondo2022temperature}. Due to the flexibility and diversity of optical field control techniques, optically controlled skyrmions have gradually become a very promising method for manipulating skyrmions\cite{berruto2018laser, yudin2017light, flovik2017generation, je2018creation, khoshlahni2019ultrafast,  vinas2022microscopic, kern2022tailoring, truc2023light, li2024room, yambe2024dynamical}. 

In recent years, research on the optical excitation and manipulation of skyrmions has become increasingly popular. Optical vortex, with their unique properties, have been widely used in the study of optically excited and manipulated skyrmions. In 2017, Fujita et al. published a study on the ultrafast generation of magnetic topological defects using vortex beams. They demonstrated that the temperature effect of a laser can influence the arrangement of magnetic moment vectors in magnetic materials, thereby printing light profiles in magnetic materials to generate magnetic topological defects \cite{fujita2017ultrafast}. In the same year, they theoretically simulated the generation of skyrmions on ferromagnetic films using LG beams \cite{fujita2017encoding}. In 2018, Yang et al. simulated the process in which the OAM carried by vortex beams causes skyrmions to move in circular orbits around the optical axis \cite{yang2018photonic}. Recently, Jiang et al. used focused light and the inverse Faraday effect to generate a stable radially polarized magnetic field, which in turn affected magnetic materials to produce skyrmions, and other magnetic spin textures \cite{jiang2024generation}. 

Despite significant breakthroughs in the field of optically controlled skyrmions, the technology is still in its early stages, and the mechanism of optical generation of skyrmions remains unclear. How to excite and manipulate magnetic skyrmions and other interesting magnetic textures through the manipulation of optical fields is a topic worthy of further research. In the optomagnetic system, it would be of great significance if optical fields could be used to control the ultrafast generation and manipulation of the number of skyrmions.

The main goal of this paper is to investigate the interaction between circularly polarized LG (CPLG) beams and chiral ferromagnets. In the study of CPLG beams, we found that the chirality of the SAM and OAM carried by the light beams can produce magnetic fields with various chirality patterns. Such vectorial magnetic fields with various chiral patterns are likely to excite different types and numbers of topological magnetic textures in magnetic materials. Moreover, ultrashort light pulses can enhance the amplitude of the  magnetic component of the optical fields to the Tesla level in a short time, achieving ultrafast interaction between the optical magnetic field and the magnetic moments in the magnetic material \cite{assouline2024helicity}. Additionally, sub-wavelength optical vortex formed by breaking the diffraction limit of terahertz beams can meet the requirements for the interaction between light and spins in magnetic materials \cite{fujita2017encoding}. 

Here, we will establish a numerical model of the optomagnetic system and perform numerical calculations based on the Landau-Lifshitz-Gilbert equation to study the ultrafast effect of the chiral magnetic field generated by ultrashort sub-wavelength vortex light pulses on magnetic materials through the Zeeman effect. This will enable the ultrafast excitation and control of the types and numbers of topological magnetic structures in magnetic materials using light.

\section{CPLG Beams }

LG beams, a solution to the Maxwell equations under the paraxial approximation, possess a helical phase structure if it has OAM. When \(m\neq0\), LG beams possess a helical phase structure and a phase singularity at \(\rho=0\). Therefore, the field distributions of LG beams vanish at \(\rho=0\), leading to the hollow intensity distribution characteristic of LG beams. The expression of LG beams in cylindrical coordinates is as follows: 

\begin{equation}
B_{m, p}(\rho, \varphi, 0) =  (\dfrac{\rho}{W})^{ \left| m \right|}e^{({-{\frac{\rho^2}{W^2}}+im\varphi })}L_p^{\left| m \right|}(\dfrac{2\rho^2}{W^2})\vec{e}_{p} ,\tag{1}
\end{equation}
where $L_{p}^{\left| m \right|}$ is the Laguerre-Gaussian polynomials, \(m\) the number of OAM, \(p\) the radial index (when \(p=0\), the beam has only one ring structure, whereas when \(p>0\), the beam exhibits multiple ring structures), and \(W\) the beam waist. \(\vec{e}_{p}\) is the polarization vector, and CPLG beams are described by \(\vec{e}_{p} = \hat{x}\pm i\hat{y}\).

CPLG beams have chiral magnetic fields. On the one hand, the SAM of left-circularly polarized Laguerre-Gaussian (LCPLG) beams and right-circularly polarized Laguerre-Gaussian (RCPLG) beams exhibit opposite chirality. The number and chirality of the OAM also vary across different types of CPLG beams. On the other hand, under the joint influence of these two types of angular momenta, the magnetic field polarization of CPLG beams undergoes rearrangement, transforming into vector beams (where the spatial distribution of polarization becomes non-uniform) and exhibiting distinct chiral patterns, thereby exerting distinct effects on chiral magnets. Here, we show some examples in FIG. 1. 

Locally, when the topological charge \(m=-1\), the magnetic field at each position rotates counterclockwise (CCW) over time due to the SAM in LCPLG beams. Globally, due to the influence of OAM, the direction of the magnetic field varies at different positions at the same time. Consequently, at certain moments, there can be two kinds of hollow helical magnetic fields with two distinct chirality (clockwise (CW) or CCW), as illustrated in FIG. 1(a) and (b). Considering that skyrmionium, a unique topological structure resembling a doughnut \cite{zhang2016control, yang2023reversible, gobel2019electrical, seng2021direct}, exhibits similar topological features to these chiral magnetic fields, it provides insight into the potential induction of skyrmionium.

When the topological charge \(m = 0\), there is no influence of OAM. The magnetic field is solely affected by SAM, with the magnetic field vector being uniform at each position and locally rotating CCW over time, as illustrated in the vector field snapshot shown in FIG. 1(f).

In contrast, when the topological charge \(m\) is neither -1 nor 0, the combined effects of SAM and OAM result in the magnetic field being divided into \(2\lvert m+1 \rvert\) regions. Therefore, there can be two kinds of chiral magnetic fields with two kinds of chirality (CW or CCW) at certain moments in each region, as depicted in FIG. 1(c), (d), and (e), which illustrate the cases where \(m=1\), generating four regions of chiral magnetic fields; \(m=2\), generating six regions; and \(m=-2\), generating two regions, respectively.

\begin{figure}
\centering
\includegraphics[width=1\linewidth]{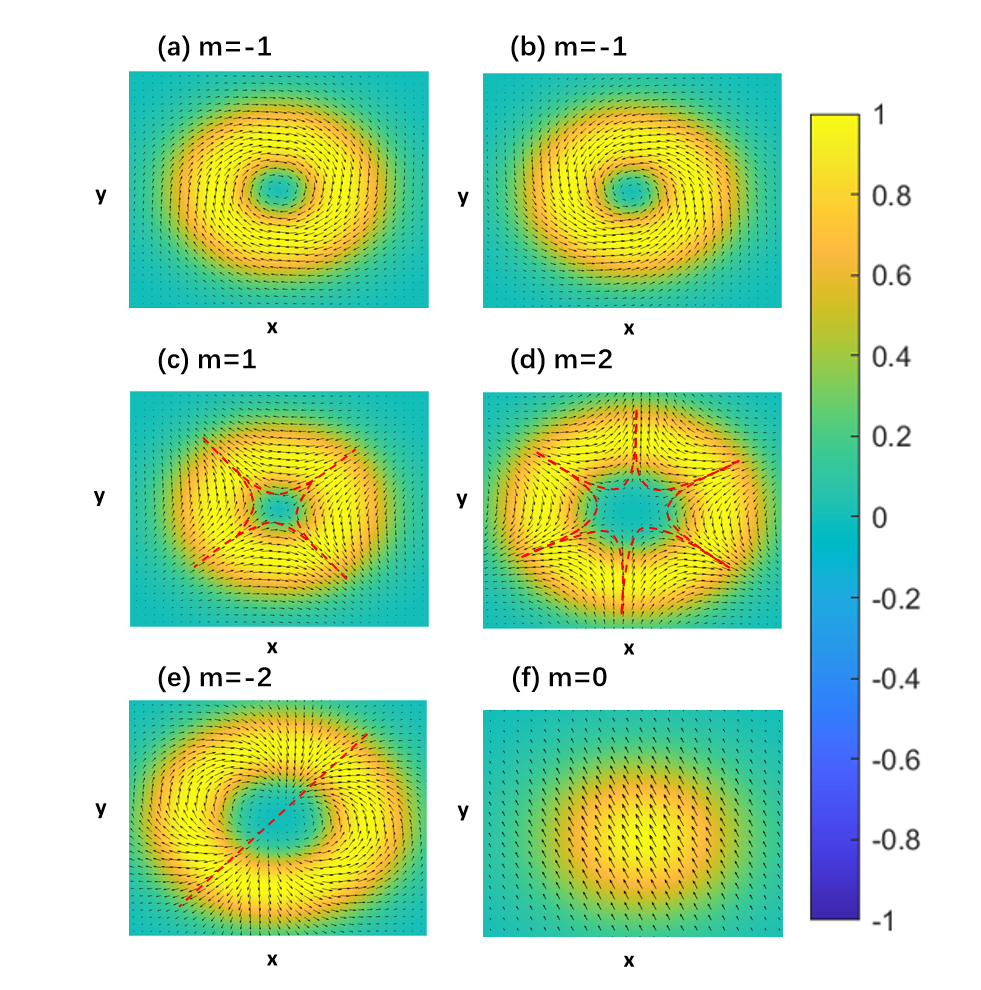}
\caption{\label{FIG1}Snapshots of chiral magnetic fields generated by the LCPLG beams: with (a) and (b) \(m=-1\) at different snapshots; (c) \(m=1\); (d) \(m=2\); (e) \(m=-2\); (f) \(m=0\).
}
\end{figure}

\section{Numerical Model}

We will verify our hypothesis through numerical simulations, examining the impact of the vectorially chiral magnetic fields from CPLG beams on chiral magnets. The model employed is shown in FIG. 2. We use CPLG beams as light source and investigate the effects on chiral magnets for different values of the topological charge
\(m= \) -1, 0, 1, 2. We simulate a chiral ferromagnet with a size of 201×201 lattices, where the initial magnetic moment vector in each lattice site is (0, 0, 1). The Hamiltonian of this chiral ferromagnetic system is:

\begin{equation}
\begin{aligned}
H = -J\sum\limits_{\vec{r}}\vec{m}_{\vec{r}} \cdot (\vec{m}_{\vec{r}+a\vec{e}_x} + \vec{m}_{\vec{r}+a\vec{e}_y}) \\
+ D\sum\limits_{\vec{r}, i}\vec{e}_i \cdot (\vec{m}_{\vec{r}} \times \vec{m}_{\vec{r}+a\vec{e}_i}) \\
- H_z\sum\limits_{\vec{r}}\vec{m}_{z} - \sum\limits_{\vec{r}}\vec{B}(\vec{r}, t) \cdot \vec{m}_{\vec{r}}. 
\end{aligned} \tag{2}
\end{equation}
It is composed of exchange, Dzyaloshinskii-Moriya interaction (DMI), and Zeeman terms. Here, \(J\) is the exchange constant, \(D\) the DMI constant, \(H_{z}\) the external constant magnetic field along the \(z\) axis, \(\vec{B}\) the magnetic field strength of the CPLG beams, \(a\) the lattice constant, and \(\vec{e}_i\) the unit vector along the \(x\) or \(y\) axis. The spin dynamics in the chiral magnet are described by the Landau-Lifshitz-Gilbert equation:

\begin{equation}
 \dfrac{d\vec{m}_{\vec{r}}}{dt}  = -\gamma \vec{m}_{\vec{r}} \times \vec{H}_{eff}+\alpha \vec{m}_{\vec{r}} \times \dfrac{d\vec{m}_{\vec{r}}}{dt} , \tag{3}
\end{equation}
where \(\gamma\) is the gyromagnetic ratio, \(\alpha\) the Gilbert damping coefficient, \(\vec{m}_{\vec{r}}\) the unit magnetic moment vector, and \(\vec{H}_{eff}=-\nabla _{\vec{m}_{\vec{r}}}(H/J)\) the effective magnetic field, which is calculated from the Hamiltonian. We use fourth-order Runge-Kutta method to calculate Eq. (3) with numerical time step \(\Delta t\) \cite{cartwright1992dynamics}.

In our simulations, we utilize a magnetic field that closely resembles the actual one, which takes the form of:

\begin{equation}
B(\vec{r}, t) = B_0 \dfrac{B_{m,p}(\vec{r})}{\max_{ \vec{\mathrm{r}}} \left|B_{m,p}(\vec{r}) \right|}e^{({-{{\frac{(t-t_0)^2}{\sigma^2}}}-{i\omega t}})} . \tag{4}
\end{equation}
Here, \(B_0\) is the coefficient of the magnetic field strength, and the quantitative change in the magnetic field strength is achieved by adjusting \(B_0\) . The modulated magnetic field strength varies with time, first increasing and then decreasing. \(t_0\) represents the moment corresponding to the maximum magnetic field strength within one period, \(\sigma\) the pulse duration, and \(\omega\) the light frequency. 

To simplify our numerical simulations, we use the simplest LG beams with \(p = 0\) and we nondimensionalize the physical quantities used in the simulation. For this simulation, we set one unit of time to \(\hbar/J\), which is 0.505 ps for \(J = 1.3\) meV and set one unit of length to 1 nm. Hereafter, we take \(\hbar=J=1\) and nondimensionalize the other constants. The relevant parameter values are assigned as follows in the simulation below: \(D=0.46\), \(\Delta t=0.05\), \(\gamma=1\), \(\alpha=0.1\), \(t_{0} = 7.72\), \(\sigma = 2.57\), \(\omega=0.285\) which is about \(0.09\) THz, lattice constant \(a=1\) which is about \(1\) nm, and \(W = 10a\). 

\begin{figure}
\centering
\includegraphics[width=0.5\linewidth]{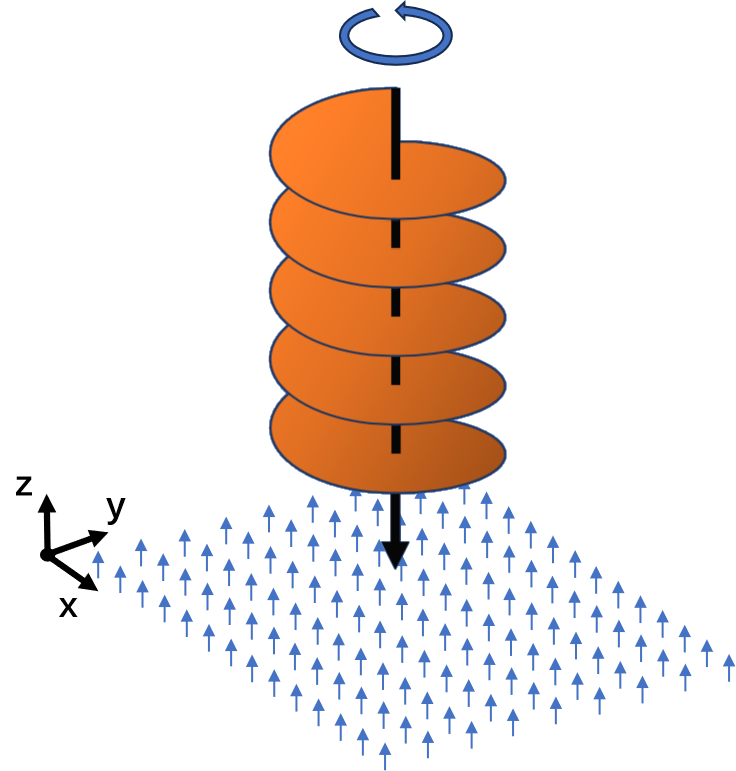}

\caption{\label{FIG2}Schematic of the model of a circularly polarized optical vortex interacting with a chiral ferromagnet.}
\end{figure}

\section{Results and discussion}

\subsection{CPLG without OAM}

First, we consider circularly polarized light beams without topological charge. In this case, the magnetic field from CPLG beams will be uniform which have no helical chirality. It seems that the conditions are not favorable for skyrmion generation. However, as long as the magnetic field strength is sufficient to flip the magnetic moments, i.e., \(B_0\) is large enough, an isolated skyrmion can still be generated under the
influence of the DMI. 
Here, we explore the interactions of LCPLG beams with two types of chiral magnets (denoted as CW and CCW to represent DMI with opposite chirality), as well as the interactions of RCPLG beams with these two chiral magnet types. We set \(H_{z} = 0.038\), and the initial spin state at \(t = 0\) is (0, 0, 1) for every lattice site. In FIG. 3, we show the result of the simulation at \(t = 80\) with these conditions.

In FIG. 3(a) and (c), we present the spin state at each lattice site in the magnetic material with CW chirality of DMI under the influence of the LCPLG beam, with \(B_0 = 0.140\) and \(B_0 = 0.180\), respectively.
In FIG. 3(b) and (d), we present the corresponding results in the magnetic material with CCW chirality of DMI under the influence of LCPLG beam, with \(B_0\) values of 0.140 and 0.180.
Finally, in FIG. 3(e-h), we present the results with the same conditions as (a-d), except that the light source was changed to RCPLG beam. 

By comparing FIG. 3(a) and (c), it can be seen that the larger the optical magnetic field strength, the larger the generated skyrmion. This is because a stronger magnetic field injects more energy into the chiral magnetic system, influencing a larger number of magnetic moments, which in turn tends to result in the formation of larger skyrmions.
Furthermore, by comparing FIG. 3(a) and (e), it is evident that the threshold magnetic field strength required for LCPLG beams to generate an isolated skyrmion is lower than that for RCPLG beams.
Additionally, a comparison between FIG. 3(a) and (b) reveals that the difference in DMI chirality (CW and CCW) leads to a 180° angular variation in the generated skyrmions, with opposite magnetic moment directions within each skyrmions.
Note that in FIG. 3(c) and (g), the variation in SAM chirality leads to a 90° angular difference in the generated skyrmions. This is due to the opposite polarization directions of the LCPLG and RCPLG beams. As the SAM chirality affects the threshold magnetic field strength for skyrmion generation, under the same magnetic field strength, the size of the skyrmions generated by LCPLG beams are larger than those generated by RCPLG beams. 

Therefore, when \(m=0\), the magnetic field vector at each position remains the same, leading to the absence of additional chiral patterns in the magnetic field. Nevertheless, the light field at this moment still possesses chirality in the degree of freedom of SAM. As a result, an isolated skyrmion can still be produced, and a larger magnetic field strength is required. The arrangement of the magnetic moments is influenced by the interplay of the chirality of SAM and DMI. 

\begin{figure*}
\includegraphics[width=1\linewidth]{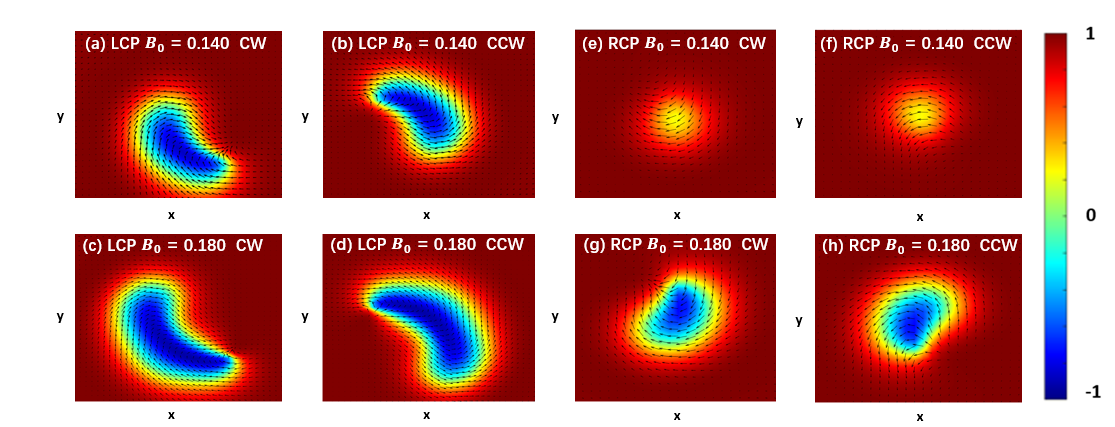}
\caption{\label{FIG3}Magnetization dynamics of the chiral magnet under the influence of CPLG beams with \(m=0\). The parameters are  \(a = 1\), \(W = 10a\), \(t_{0} = 7.72\), \(\sigma = 2.57\), \(J = 1\), \(D = 0.46\), \(H_{z} = 0.038\), \(\omega = 0.285\) and \(t = 80\). (a-b) Snapshot of spin states for LCPLG and \(B_0 = 0.140\) in (a) CW, and (b) CCW chiral magnet; (c-d) Snapshot of spin states for LCPLG and \(B_0 = 0.180\) in (c) CW, and (d) chiral magnet; (e-h) Under the same conditions as (a-d) except that the light source was changed to RCPLG beam. }
\end{figure*}

\subsection{CPLG with m = -1}

Next, we are in a position to consider circularly polarized vortex beams. When \(m = -1\), the magnetic field from LCPLG exhibits hollow helical chirality, which is very similar to the topological structure of skyrmionium. 
Therefore, we simulate the magnetization dynamics of two types of chiral magnetic materials (CW and CCW) under different magnetic field strengths of the LCPLG beams, with \(H_{z} = 0.1\), and present the results at \(t = 80\) shown in FIG. 4. 

First, we discuss the results when the LCPLG beams affect the chiral magnetic material with CW chirality, as shown in FIG. 4 (a) and (c). 
FIG. 4 (a) shows the spin state for each lattice site in the magnetic material with CW chirality of DMI under the influence of LCPLG beam with \(B_0 = 0.160\). The magnetic field strength is not sufficient to produce skyrmionium. However, we observe that a ring-shaped region appeared in the chiral magnetic material, where most of the magnetic moments lie in the plane (green region), and there are four small regions where the magnetic moments have flipped (blue region). 
In FIG. 4 (b), we increase the magnetic field strength to, for example, \(B_0 = 0.185\). The blue region expands to the entire ring-shaped region, forming a skyrmionium with CW chirality, consistent with the chirality (CW) of the chiral magnetic material. 
In FIG. 4 (c), we show the phase diagram in the interval of \(B_0\) from 0.070 to 0.190. As \(B_0\) increases, a ring-shaped region with magnetic moments lying in the plane is first generated in the magnetic material, and then the magnetic moments in the ring-shaped region flip from small regions to the entire ring-shaped region, eventually forming skyrmionium.

Second, we discuss the cases when LCPLG beams affect the chiral magnetic material with CCW chirality, as shown in Fig. 4 (d) and (e). FIG. 4 (d) shows the spin state for each lattice site in the magnetic material with CCW chirality of DMI under the influence of LCPLG beam with \(B_0 = 0.072\). An isolated skyrmion is generated in the magnetic material, rather than skyrmionium. The reason is that the magnetic field strength \(B_0\) is low such that the magnetic moments affected by LCPLG beams flip relatively slowly. Although the magnetic field generated by the LCPLG beams is hollow, with the magnetic moments in the central region unaffected by the field, these moments will still gradually flip due to exchange interaction and DMI, eventually leading to the formation of a skyrmion. 
In FIG. 4 (e), we present the result when \(B_0 = 0.073\). With an increased magnetic field strength \(B_0\), the disparity in magnetic field intensity between the region influenced by the magnetic field and the central region (which remains unaffected) becomes more pronounced.  The magnetic moments affected by the magnetic field flip faster, causing the others not affected by the magnetic field to be unable to be driven to flip, thus generating a skyrmionium with CCW chirality, consistent with the chirality (CCW) of the magnetic material. 
FIG. 4 (f) presents the phase diagram in the interval of \(B_0\) from 0.070 to 0.190. It reveals that an isolated skyrmion is first generated in the magnetic material, and as \(B_0\) increases, the center of the skyrmion, with unflipped magnetic moments, evolves into a skyrmionium. When \(B_0\) continues to increase, skyrmionium is generated with a larger unflipped central region.

Therefore, the magnetic field from LCPLG beams with \(m = -1\) can induce skyrmionium with CW or CCW chirality in two different DMI chiral magnetic materials. When the chirality of DMI is CCW, since its chirality is the same as the SAM chirality of LCPLG beams, this chiral magnetic material can generate skyrmionium at a smaller magnetic field strength. When the magnetic field strength is sufficiently small, skyrmions can still be generated. However, the other type of DMI chirality (CW) magnetic material requires a larger magnetic field strength to form skyrmionium.

\begin{figure*}
\includegraphics[width=1\linewidth]{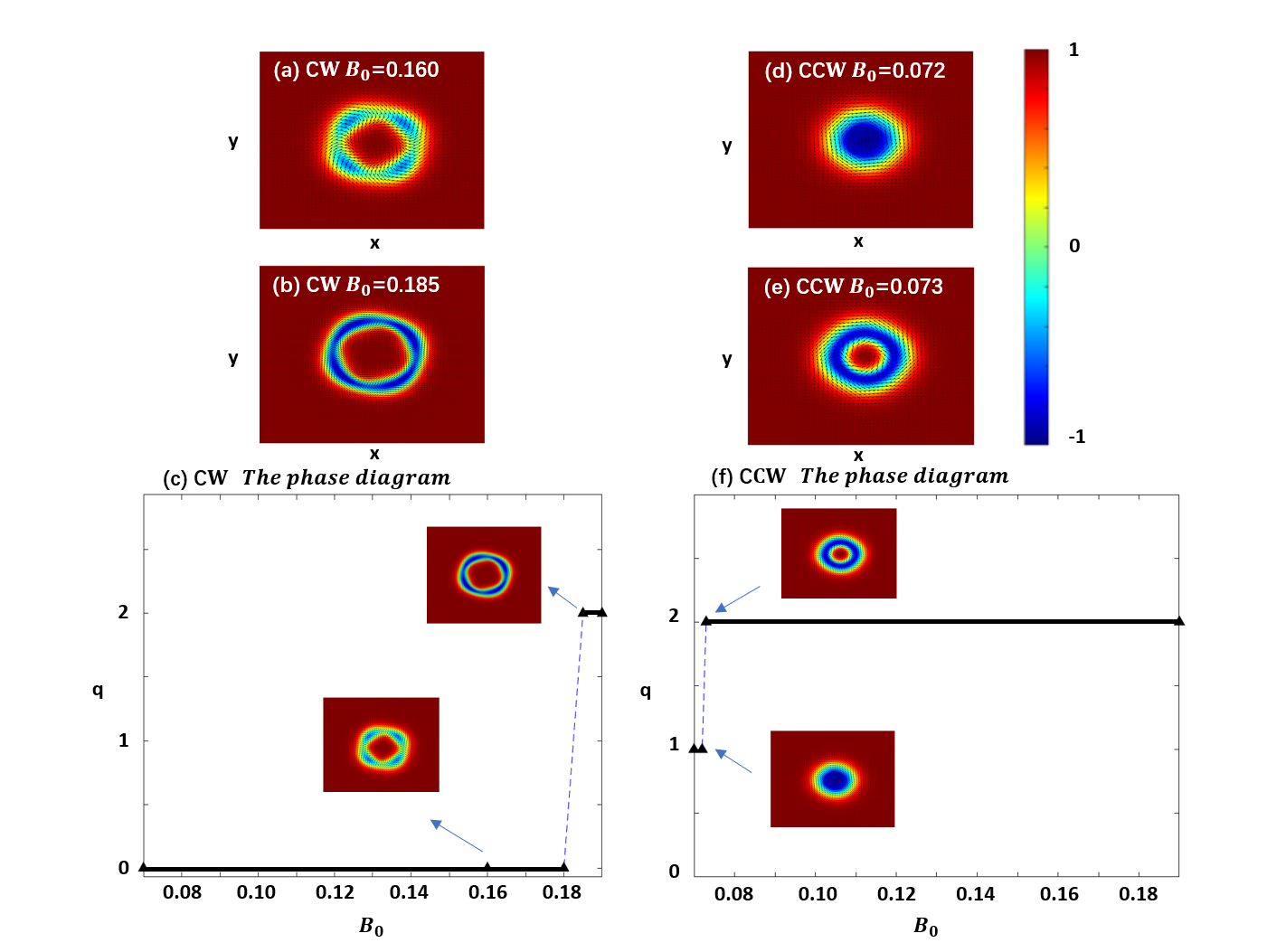}
\caption{\label{FIG4}Magnetization dynamics of the chiral magnet under the influence of LCPLG beams with \(m=-1\). The parameters are  \(a = 1\), \(W = 10a\), \(t_{0} = 7.72\), \(\sigma = 2.57\), \(J = 1\), \(D = 0.46\), \(H_{z} = 0.1\), \(\omega = 0.285\) and \(t = 80\). (a-b) Snapshot of spin states for (a) \(B_0 = 0.160\), and (b) \(B_0 = 0.185\) in CW chiral magnet; (c) Phase diagram for \(B_0 = 0.070-0.190\) in CW chiral magnet; (d-e) Snapshot of spin states for (d) \(B_0 = 0.072\), and (e) \(B_0 = 0.073\) in CCW chiral magnet; (f) Phase diagram for \(B_0 = 0.070-0.190\) in CCW chiral magnet. }
\end{figure*}

\subsection{CPLG with m = 1, 2}

Finally, we consider circularly polarized vortex light beams with topological charge \(m = 1, 2\). For one thing, LCPLG beams carry SAM, with the direction of the magnetic field on each position rotating CCW over time. For another, the chiral magnetic field from LCPLG beams can be globally divided into \(2\lvert m+1 \rvert\) regions, each exhibiting either CW or CCW chirality, as already shown in FIG. 1. This will be conducive to inducing multiple skyrmions in the magnet, and the arrangement of magnetic moments in the skyrmions will be different due to the chirality (CW or CCW) of the chiral magnet. We simulate the effect of LCPLG beams with \(m = 1, 2\) on a magnet with CW chirality, using the same parameters as those for CPLG in the previous subsection.

For \(m = 1\), again we observe that the interplay of the SAM and OAM of the LCPLG beams is reflected in the formation of a magnetic field with four chiral regions, giving a hint on the induction of up to four skyrmions. 
As shown in FIG. 5 (a), when the magnetic field strength \(B_0= 0.110\), in the early stage of the magnetic field's influence, a ring-shaped region forms in the magnet, where the magnetic moments are flipped.
Due to the relatively low magnetic field strength, this ring-shaped region can only split into two flipped regions under the action of the LCPLG. The magnetic moments in these two regions evolve over time, eventually forming two skyrmions with circular shapes \cite{fujita2017encoding}. 
When the magnetic field strength increases to 0.150 (FIG. 5 (b)), two strip-shaped skyrmions appear in the magnet. This occurs because the ring-shaped region formed in the early stage of the magnetic field's action splits into four flipped regions under a stronger magnetic field. However, since the distance between the flipped regions is relatively small, pairwise adhesion takes place, ultimately resulting in the formation of two strip-shaped skyrmions.
Further, as the magnetic field strength increases to 0.153 (FIG. 5 (c)), the ring-shaped flipped region continues to split into four distinct flipped regions. However, the distance between these regions is large enough to prevent pairwise adhesion. This results in the formation of four distinct skyrmions. Finally, the phase diagram in FIG. 5 (d) illustrates the number of skyrmions generated in the magnet as a function of the magnetic field strength \(B_0\), ranging from 0.110 to 0.160. Initially, the number of skyrmions is two. As the strength of the magnetic field increases, the skyrmions will transform into a stripe-like shape, but there will still be two of them. In contrast, when the magnetic field strength is even higher, four skyrmions will be generated, which verifies our conjecture: the magnetic field generated by LCPLG beams with \(m = 1\) can induce up to four skyrmions in the magnet.

For \(m = 2\), the magnetic field produced by LG light features six chiral regions, suggesting that up to six skyrmions may be generated.
As shown in FIG. 5 (e), when the magnetic field strength \(B_0\) = 0.160, three skyrmions will be generated in the magnet. Although, in the early stage of the magnetic field's application, the ring-shaped flipped region splits into six flipped regions, the relatively small distance between these clustered regions causes them to pair up and merge into three flipped regions. This process ultimately leads to the formation of three skyrmions \cite{fujita2017encoding}. 
When \(B_0\) = 0.170, as seen in FIG. 5 (h), the distance between each generated region is large enough to prevent adhesion, and six skyrmions can eventually be formed. 
However, in the range of \(B_0\) = 0.160 to 0.165, the distance between the six flipped regions is not sufficient to avoid adhesion, causing some regions to cluster together, which results in the formation of a different number of skyrmions.
\begin{figure*}
\includegraphics[width=1\linewidth]{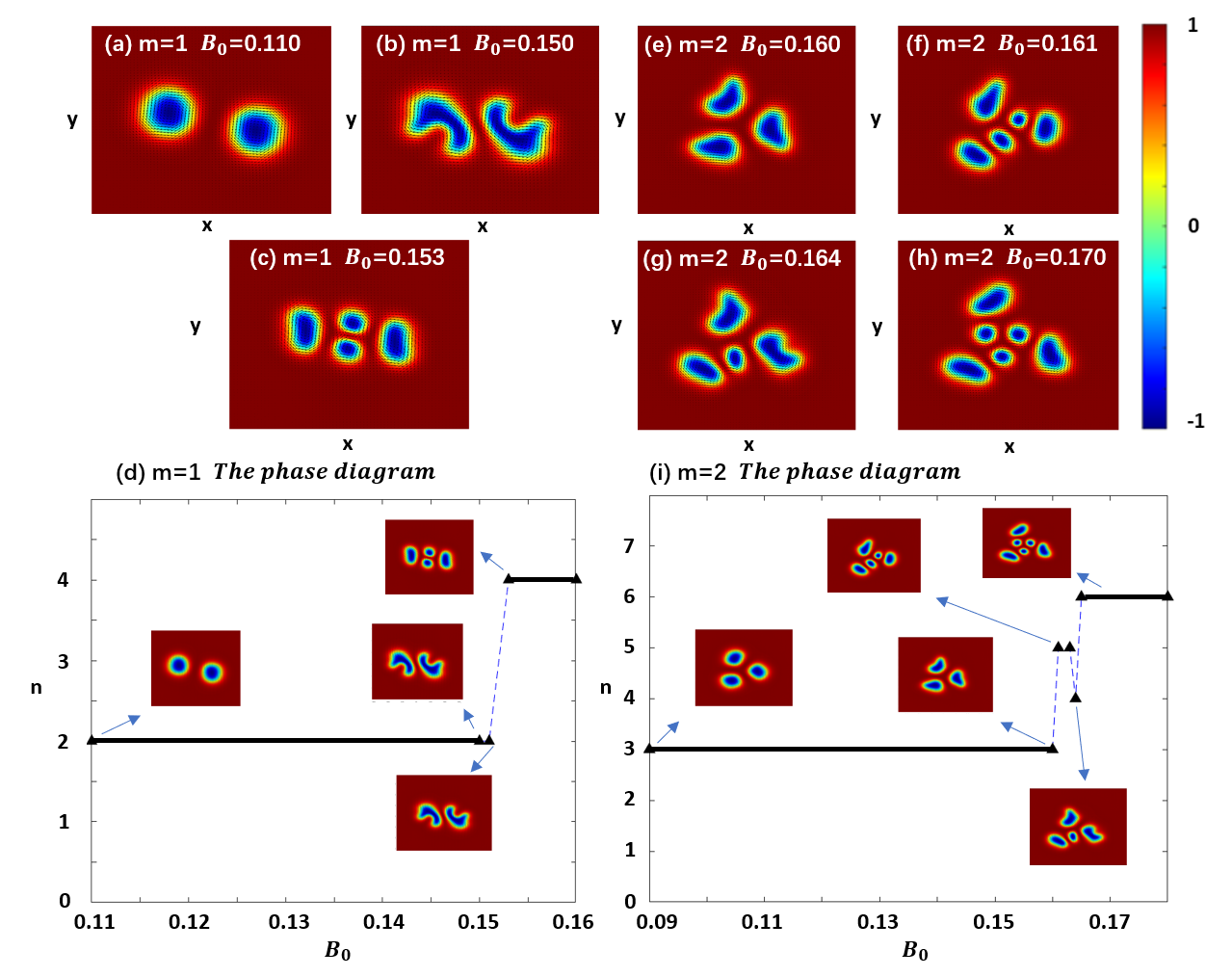}
\caption{\label{FIG5}Magnetization dynamics of CW chiral magnet under the influence of LCPLG beams with \(m=1,2\). The parameters are  \(a = 1\), \(W = 10a\), \(t_{0} = 7.72\), \(\sigma = 2.57\), \(J = 1\), \(D = 0.46\), \(H_{z} = 0.1\), \(\omega = 0.285\) and \(t = 80\). (a-c) Snapshot of spin states for \(m = 1\) and (a) \(B_0 = 0.110\), (b) \(B_0 = 0.150\), and (c) \(B_0 = 0.153\); (d) Phase diagram for \(m = 1\) and \(B_0 = 0.110-0.160\); (e-h) Snapshot of spin states \(m = 2\) and (e) \(B_0 = 0.160\), (f) \(B_0 = 0.161\), (g) \(B_0 = 0.164\), and (h) \(B_0 = 0.170\); (i) Phase diagram for \(m = 2\) and \(B_0 = 0.090-0.180\). }
\end{figure*}
For example, in FIG.5 (f), when \(B_0\) = 0.161, five skyrmions are generated, and in FIG. 5 (g), when \(B_0\) = 0.164, four skyrmions are generated. In our simulations, we observed that the number of generated skyrmions falls within the range of 3 to 6. 
In FIG. 5 (i), the phase diagram illustrates the number of skyrmions as a function of the magnetic field strength \(B_0\) in the range of 0.090 to 0.180. When the magnetic field strength is low, the ring-shaped flipped region generated in the early stage of the magnetic field's action splits into three flipped regions, ultimately resulting in the formation of three skyrmions. When the magnetic field strength is strong, six flipped regions initially form. However, due to pairwise adhesion, only three skyrmions are ultimately generated.
In the range of \(B_0\) = 0.160 to 0.165, due to the potential binding between skyrmions, dashed lines are employed to indicate that the number of skyrmions may exceed three, but remain below six. 
After \(B_0\) = 0.165, the magnet, influenced by the chiral magnetic fields, can stably form six skyrmions.

Therefore, we find that when \(m\neq0, -1\), the magnetic field generated by LCPLG beams acting on chiral magnets can produce at least \(\lvert m+1 \rvert\) skyrmions \cite{fujita2017encoding} and at most \(2\lvert m+1 \rvert\) skyrmions, and the number of skyrmions generated can be controlled by adjusting the magnetic field strength.

\section{Conclusion}

In summary, we uncover the mechanism by which vortex light with circular polarization generates skyrmions or skyrmionium and determine the number of skyrmions produced, when CPLG beams are used to manipulate a chiral magnet. CPLG beams can generate chiral magnetic fields by the interplay of SAM and OAM, i.e., the spatial and polarization degrees of freedom. These chiral magnetic fields generated by CPLG beams can induce the formation of an isolated skyrmionium or skyrmions which have the same chirality as DMI in chiral magnetic materials. An isolated skyrmion can be generated while we use CPLG without OAM and a single skyrmionium can be generated while we use LCPLG with \(m=-1\). In addition, we can generate \(\lvert m+1 \rvert\) or \(2\lvert m+1 \rvert\) skyrmions using LCPLG with \(m\neq 0,-1\). Under certain conditions of magnetic field strength, the distance between the flipped regions may not be large enough, causing some skyrmions to merge into one. In these cases, the number of skyrmions generated can range from \(\lvert m+1 \rvert\) to \(2\lvert m+1 \rvert\). This is advantageous for achieving skyrmion-based encoding through modulation of light intensity, thereby guiding the design of skyrmion-based memory devices. Thinking forward, we believe that the manipulation and optimization of light fields \cite{chen2021engineering} can be used to control the distance between skyrmions. In this way, we can control the clustering of specific skyrmions, thereby stably regulating the number of skyrmions within the range of \(\lvert m+1 \rvert\) to \(2\lvert m+1 \rvert\). Hence, our work may pave the way for the ultrafast design of skyrmion-based encoding and memory devices. 

\section{Acknowledgment}

This work was supported by National Natural Science Foundation of China (Project No. 12104296) and Natural Science Foundation of Guangdong (Grant No. 2023A1515011345).

\nocite{*}

\bibliography{apssamp}

\begin{thebibliography}{68}%
\makeatletter
\providecommand \@ifxundefined [1]{%
 \@ifx{#1\undefined}
}%
\providecommand \@ifnum [1]{%
 \ifnum #1\expandafter \@firstoftwo
 \else \expandafter \@secondoftwo
 \fi
}%
\providecommand \@ifx [1]{%
 \ifx #1\expandafter \@firstoftwo
 \else \expandafter \@secondoftwo
 \fi
}%
\providecommand \natexlab [1]{#1}%
\providecommand \enquote  [1]{``#1''}%
\providecommand \bibnamefont  [1]{#1}%
\providecommand \bibfnamefont [1]{#1}%
\providecommand \citenamefont [1]{#1}%
\providecommand \href@noop [0]{\@secondoftwo}%
\providecommand \href [0]{\begingroup \@sanitize@url \@href}%
\providecommand \@href[1]{\@@startlink{#1}\@@href}%
\providecommand \@@href[1]{\endgroup#1\@@endlink}%
\providecommand \@sanitize@url [0]{\catcode `\\12\catcode `\$12\catcode `\&12\catcode `\#12\catcode `\^12\catcode `\_12\catcode `\%12\relax}%
\providecommand \@@startlink[1]{}%
\providecommand \@@endlink[0]{}%
\providecommand \url  [0]{\begingroup\@sanitize@url \@url }%
\providecommand \@url [1]{\endgroup\@href {#1}{\urlprefix }}%
\providecommand \urlprefix  [0]{URL }%
\providecommand \Eprint [0]{\href }%
\providecommand \doibase [0]{https://doi.org/}%
\providecommand \selectlanguage [0]{\@gobble}%
\providecommand \bibinfo  [0]{\@secondoftwo}%
\providecommand \bibfield  [0]{\@secondoftwo}%
\providecommand \translation [1]{[#1]}%
\providecommand \BibitemOpen [0]{}%
\providecommand \bibitemStop [0]{}%
\providecommand \bibitemNoStop [0]{.\EOS\space}%
\providecommand \EOS [0]{\spacefactor3000\relax}%
\providecommand \BibitemShut  [1]{\csname bibitem#1\endcsname}%
\let\auto@bib@innerbib\@empty
\bibitem [{\citenamefont {Au}\ \emph {et~al.}(2013)\citenamefont {Au}, \citenamefont {Dvornik}, \citenamefont {Davison}, \citenamefont {Ahmad}, \citenamefont {Keatley}, \citenamefont {Vansteenkiste}, \citenamefont {Van~Waeyenberge},\ and\ \citenamefont {Kruglyak}}]{au2013direct}%
  \BibitemOpen
  \bibfield  {author} {\bibinfo {author} {\bibfnamefont {Y.}~\bibnamefont {Au}}, \bibinfo {author} {\bibfnamefont {M.}~\bibnamefont {Dvornik}}, \bibinfo {author} {\bibfnamefont {T.}~\bibnamefont {Davison}}, \bibinfo {author} {\bibfnamefont {E.}~\bibnamefont {Ahmad}}, \bibinfo {author} {\bibfnamefont {P.~S.}\ \bibnamefont {Keatley}}, \bibinfo {author} {\bibfnamefont {A.}~\bibnamefont {Vansteenkiste}}, \bibinfo {author} {\bibfnamefont {B.}~\bibnamefont {Van~Waeyenberge}},\ and\ \bibinfo {author} {\bibfnamefont {V.}~\bibnamefont {Kruglyak}},\ }\bibfield  {title} {\bibinfo {title} {Direct excitation of propagating spin waves by focused ultrashort optical pulses},\ }\href@noop {} {\bibfield  {journal} {\bibinfo  {journal} {Physical Review Letters}\ }\textbf {\bibinfo {volume} {110}},\ \bibinfo {pages} {097201} (\bibinfo {year} {2013})}\BibitemShut {NoStop}%
\bibitem [{\citenamefont {Blanco}\ \emph {et~al.}(2018)\citenamefont {Blanco}, \citenamefont {Cambronero}, \citenamefont {Flores-Arias}, \citenamefont {Conejero~Jarque}, \citenamefont {Plaja},\ and\ \citenamefont {Hern{\'a}ndez-Garc{\'\i}a}}]{blanco2018ultraintense}%
  \BibitemOpen
  \bibfield  {author} {\bibinfo {author} {\bibfnamefont {M.}~\bibnamefont {Blanco}}, \bibinfo {author} {\bibfnamefont {F.}~\bibnamefont {Cambronero}}, \bibinfo {author} {\bibfnamefont {M.~T.}\ \bibnamefont {Flores-Arias}}, \bibinfo {author} {\bibfnamefont {E.}~\bibnamefont {Conejero~Jarque}}, \bibinfo {author} {\bibfnamefont {L.}~\bibnamefont {Plaja}},\ and\ \bibinfo {author} {\bibfnamefont {C.}~\bibnamefont {Hern{\'a}ndez-Garc{\'\i}a}},\ }\bibfield  {title} {\bibinfo {title} {Ultraintense femtosecond magnetic nanoprobes induced by azimuthally polarized laser beams},\ }\href@noop {} {\bibfield  {journal} {\bibinfo  {journal} {ACS Photonics}\ }\textbf {\bibinfo {volume} {6}},\ \bibinfo {pages} {38} (\bibinfo {year} {2018})}\BibitemShut {NoStop}%
\bibitem [{\citenamefont {Fujita}\ \emph {et~al.}(2019)\citenamefont {Fujita}, \citenamefont {Tada},\ and\ \citenamefont {Sato}}]{fujita2019accessing}%
  \BibitemOpen
  \bibfield  {author} {\bibinfo {author} {\bibfnamefont {H.}~\bibnamefont {Fujita}}, \bibinfo {author} {\bibfnamefont {Y.}~\bibnamefont {Tada}},\ and\ \bibinfo {author} {\bibfnamefont {M.}~\bibnamefont {Sato}},\ }\bibfield  {title} {\bibinfo {title} {Accessing electromagnetic properties of matter with cylindrical vector beams},\ }\href@noop {} {\bibfield  {journal} {\bibinfo  {journal} {New Journal of Physics}\ }\textbf {\bibinfo {volume} {21}},\ \bibinfo {pages} {073010} (\bibinfo {year} {2019})}\BibitemShut {NoStop}%
\bibitem [{\citenamefont {Lin}\ \emph {et~al.}(2022)\citenamefont {Lin}, \citenamefont {Wang}, \citenamefont {Yuan},\ and\ \citenamefont {Chen}}]{lin2022all}%
  \BibitemOpen
  \bibfield  {author} {\bibinfo {author} {\bibfnamefont {S.}~\bibnamefont {Lin}}, \bibinfo {author} {\bibfnamefont {L.}~\bibnamefont {Wang}}, \bibinfo {author} {\bibfnamefont {L.}~\bibnamefont {Yuan}},\ and\ \bibinfo {author} {\bibfnamefont {X.}~\bibnamefont {Chen}},\ }\bibfield  {title} {\bibinfo {title} {All-optical control of the photonic hall lattice in a pumped waveguide array},\ }\href@noop {} {\bibfield  {journal} {\bibinfo  {journal} {Physical Review Applied}\ }\textbf {\bibinfo {volume} {17}},\ \bibinfo {pages} {064029} (\bibinfo {year} {2022})}\BibitemShut {NoStop}%
\bibitem [{\citenamefont {Fu}\ \emph {et~al.}(2020)\citenamefont {Fu}, \citenamefont {Wang}, \citenamefont {Huang}, \citenamefont {Kartashov}, \citenamefont {Torner}, \citenamefont {Konotop},\ and\ \citenamefont {Ye}}]{fu2020optical}%
  \BibitemOpen
  \bibfield  {author} {\bibinfo {author} {\bibfnamefont {Q.}~\bibnamefont {Fu}}, \bibinfo {author} {\bibfnamefont {P.}~\bibnamefont {Wang}}, \bibinfo {author} {\bibfnamefont {C.}~\bibnamefont {Huang}}, \bibinfo {author} {\bibfnamefont {Y.~V.}\ \bibnamefont {Kartashov}}, \bibinfo {author} {\bibfnamefont {L.}~\bibnamefont {Torner}}, \bibinfo {author} {\bibfnamefont {V.~V.}\ \bibnamefont {Konotop}},\ and\ \bibinfo {author} {\bibfnamefont {F.}~\bibnamefont {Ye}},\ }\bibfield  {title} {\bibinfo {title} {Optical soliton formation controlled by angle twisting in photonic moir{\'e} lattices},\ }\href@noop {} {\bibfield  {journal} {\bibinfo  {journal} {Nature Photonics}\ }\textbf {\bibinfo {volume} {14}},\ \bibinfo {pages} {663} (\bibinfo {year} {2020})}\BibitemShut {NoStop}%
\bibitem [{\citenamefont {Graf}\ \emph {et~al.}(2021)\citenamefont {Graf}, \citenamefont {Sharma}, \citenamefont {Huebl},\ and\ \citenamefont {Kusminskiy}}]{graf2021design}%
  \BibitemOpen
  \bibfield  {author} {\bibinfo {author} {\bibfnamefont {J.}~\bibnamefont {Graf}}, \bibinfo {author} {\bibfnamefont {S.}~\bibnamefont {Sharma}}, \bibinfo {author} {\bibfnamefont {H.}~\bibnamefont {Huebl}},\ and\ \bibinfo {author} {\bibfnamefont {S.~V.}\ \bibnamefont {Kusminskiy}},\ }\bibfield  {title} {\bibinfo {title} {Design of an optomagnonic crystal: Towards optimal magnon-photon mode matching at the microscale},\ }\href@noop {} {\bibfield  {journal} {\bibinfo  {journal} {Physical Review Research}\ }\textbf {\bibinfo {volume} {3}},\ \bibinfo {pages} {013277} (\bibinfo {year} {2021})}\BibitemShut {NoStop}%
\bibitem [{\citenamefont {Lin}\ \emph {et~al.}(2023)\citenamefont {Lin}, \citenamefont {Liang}, \citenamefont {Zhang}, \citenamefont {Chen},\ and\ \citenamefont {Tsai}}]{lin2023controllable}%
  \BibitemOpen
  \bibfield  {author} {\bibinfo {author} {\bibfnamefont {S.}~\bibnamefont {Lin}}, \bibinfo {author} {\bibfnamefont {Y.}~\bibnamefont {Liang}}, \bibinfo {author} {\bibfnamefont {J.}~\bibnamefont {Zhang}}, \bibinfo {author} {\bibfnamefont {M.~K.}\ \bibnamefont {Chen}},\ and\ \bibinfo {author} {\bibfnamefont {D.~P.}\ \bibnamefont {Tsai}},\ }\bibfield  {title} {\bibinfo {title} {Controllable flatbands via non-hermiticity},\ }\href@noop {} {\bibfield  {journal} {\bibinfo  {journal} {Applied Physics Letters}\ }\textbf {\bibinfo {volume} {123}} (\bibinfo {year} {2023})}\BibitemShut {NoStop}%
\bibitem [{\citenamefont {Liang}\ \emph {et~al.}(2024)\citenamefont {Liang}, \citenamefont {Tsai},\ and\ \citenamefont {Kivshar}}]{liang2024local}%
  \BibitemOpen
  \bibfield  {author} {\bibinfo {author} {\bibfnamefont {Y.}~\bibnamefont {Liang}}, \bibinfo {author} {\bibfnamefont {D.~P.}\ \bibnamefont {Tsai}},\ and\ \bibinfo {author} {\bibfnamefont {Y.}~\bibnamefont {Kivshar}},\ }\bibfield  {title} {\bibinfo {title} {From local to nonlocal high-q plasmonic metasurfaces},\ }\href@noop {} {\bibfield  {journal} {\bibinfo  {journal} {Physical Review Letters}\ }\textbf {\bibinfo {volume} {133}},\ \bibinfo {pages} {053801} (\bibinfo {year} {2024})}\BibitemShut {NoStop}%
\bibitem [{\citenamefont {Beth}(1936)}]{beth1936mechanical}%
  \BibitemOpen
  \bibfield  {author} {\bibinfo {author} {\bibfnamefont {R.~A.}\ \bibnamefont {Beth}},\ }\bibfield  {title} {\bibinfo {title} {Mechanical detection and measurement of the angular momentum of light},\ }\href@noop {} {\bibfield  {journal} {\bibinfo  {journal} {Physical Review}\ }\textbf {\bibinfo {volume} {50}},\ \bibinfo {pages} {115} (\bibinfo {year} {1936})}\BibitemShut {NoStop}%
\bibitem [{\citenamefont {Allen}\ \emph {et~al.}(1992)\citenamefont {Allen}, \citenamefont {Beijersbergen}, \citenamefont {Spreeuw},\ and\ \citenamefont {Woerdman}}]{allen1992orbital}%
  \BibitemOpen
  \bibfield  {author} {\bibinfo {author} {\bibfnamefont {L.}~\bibnamefont {Allen}}, \bibinfo {author} {\bibfnamefont {M.~W.}\ \bibnamefont {Beijersbergen}}, \bibinfo {author} {\bibfnamefont {R.}~\bibnamefont {Spreeuw}},\ and\ \bibinfo {author} {\bibfnamefont {J.}~\bibnamefont {Woerdman}},\ }\bibfield  {title} {\bibinfo {title} {Orbital angular momentum of light and the transformation of laguerre-gaussian laser modes},\ }\href@noop {} {\bibfield  {journal} {\bibinfo  {journal} {Physical Review A}\ }\textbf {\bibinfo {volume} {45}},\ \bibinfo {pages} {8185} (\bibinfo {year} {1992})}\BibitemShut {NoStop}%
\bibitem [{\citenamefont {Lin}\ \emph {et~al.}(2019)\citenamefont {Lin}, \citenamefont {Nie}, \citenamefont {Yan}, \citenamefont {Liang}, \citenamefont {Lin}, \citenamefont {Zhao},\ and\ \citenamefont {Jia}}]{lin2019all}%
  \BibitemOpen
  \bibfield  {author} {\bibinfo {author} {\bibfnamefont {S.}~\bibnamefont {Lin}}, \bibinfo {author} {\bibfnamefont {Z.}~\bibnamefont {Nie}}, \bibinfo {author} {\bibfnamefont {W.}~\bibnamefont {Yan}}, \bibinfo {author} {\bibfnamefont {Y.}~\bibnamefont {Liang}}, \bibinfo {author} {\bibfnamefont {H.}~\bibnamefont {Lin}}, \bibinfo {author} {\bibfnamefont {Q.}~\bibnamefont {Zhao}},\ and\ \bibinfo {author} {\bibfnamefont {B.}~\bibnamefont {Jia}},\ }\bibfield  {title} {\bibinfo {title} {All-optical vectorial control of multistate magnetization through anisotropy-mediated spin-orbit coupling},\ }\href@noop {} {\bibfield  {journal} {\bibinfo  {journal} {Nanophotonics}\ }\textbf {\bibinfo {volume} {8}},\ \bibinfo {pages} {2177} (\bibinfo {year} {2019})}\BibitemShut {NoStop}%
\bibitem [{\citenamefont {Nakata}\ \emph {et~al.}(2019)\citenamefont {Nakata}, \citenamefont {Kim},\ and\ \citenamefont {Takayoshi}}]{nakata2019laser}%
  \BibitemOpen
  \bibfield  {author} {\bibinfo {author} {\bibfnamefont {K.}~\bibnamefont {Nakata}}, \bibinfo {author} {\bibfnamefont {S.~K.}\ \bibnamefont {Kim}},\ and\ \bibinfo {author} {\bibfnamefont {S.}~\bibnamefont {Takayoshi}},\ }\bibfield  {title} {\bibinfo {title} {Laser control of magnonic topological phases in antiferromagnets},\ }\href@noop {} {\bibfield  {journal} {\bibinfo  {journal} {Physical Review B}\ }\textbf {\bibinfo {volume} {100}},\ \bibinfo {pages} {014421} (\bibinfo {year} {2019})}\BibitemShut {NoStop}%
\bibitem [{\citenamefont {Banerjee}\ \emph {et~al.}(2022)\citenamefont {Banerjee}, \citenamefont {Kumar},\ and\ \citenamefont {Lin}}]{banerjee2022inverse}%
  \BibitemOpen
  \bibfield  {author} {\bibinfo {author} {\bibfnamefont {S.}~\bibnamefont {Banerjee}}, \bibinfo {author} {\bibfnamefont {U.}~\bibnamefont {Kumar}},\ and\ \bibinfo {author} {\bibfnamefont {S.-Z.}\ \bibnamefont {Lin}},\ }\bibfield  {title} {\bibinfo {title} {Inverse faraday effect in mott insulators},\ }\href@noop {} {\bibfield  {journal} {\bibinfo  {journal} {Physical Review B}\ }\textbf {\bibinfo {volume} {105}},\ \bibinfo {pages} {L180414} (\bibinfo {year} {2022})}\BibitemShut {NoStop}%
\bibitem [{\citenamefont {Guan}\ \emph {et~al.}(2023)\citenamefont {Guan}, \citenamefont {Liu}, \citenamefont {Hou}, \citenamefont {Chen}, \citenamefont {Fan}, \citenamefont {Zeng}, \citenamefont {Lu}, \citenamefont {Gao}, \citenamefont {Qin},\ and\ \citenamefont {Liu}}]{guan2023optically}%
  \BibitemOpen
  \bibfield  {author} {\bibinfo {author} {\bibfnamefont {S.}~\bibnamefont {Guan}}, \bibinfo {author} {\bibfnamefont {Y.}~\bibnamefont {Liu}}, \bibinfo {author} {\bibfnamefont {Z.}~\bibnamefont {Hou}}, \bibinfo {author} {\bibfnamefont {D.}~\bibnamefont {Chen}}, \bibinfo {author} {\bibfnamefont {Z.}~\bibnamefont {Fan}}, \bibinfo {author} {\bibfnamefont {M.}~\bibnamefont {Zeng}}, \bibinfo {author} {\bibfnamefont {X.}~\bibnamefont {Lu}}, \bibinfo {author} {\bibfnamefont {X.}~\bibnamefont {Gao}}, \bibinfo {author} {\bibfnamefont {M.}~\bibnamefont {Qin}},\ and\ \bibinfo {author} {\bibfnamefont {J.-M.}\ \bibnamefont {Liu}},\ }\bibfield  {title} {\bibinfo {title} {Optically controlled ultrafast dynamics of skyrmion in antiferromagnets},\ }\href@noop {} {\bibfield  {journal} {\bibinfo  {journal} {Physical Review B}\ }\textbf {\bibinfo {volume} {107}},\ \bibinfo {pages} {214429} (\bibinfo {year} {2023})}\BibitemShut {NoStop}%
\bibitem [{\citenamefont {Ghosh}\ \emph {et~al.}(2023)\citenamefont {Ghosh}, \citenamefont {Bl{\"u}gel},\ and\ \citenamefont {Mokrousov}}]{ghosh2023ultrafast}%
  \BibitemOpen
  \bibfield  {author} {\bibinfo {author} {\bibfnamefont {S.}~\bibnamefont {Ghosh}}, \bibinfo {author} {\bibfnamefont {S.}~\bibnamefont {Bl{\"u}gel}},\ and\ \bibinfo {author} {\bibfnamefont {Y.}~\bibnamefont {Mokrousov}},\ }\bibfield  {title} {\bibinfo {title} {Ultrafast optical generation of antiferromagnetic meron-antimeron pairs with conservation of topological charge},\ }\href@noop {} {\bibfield  {journal} {\bibinfo  {journal} {Physical Review Research}\ }\textbf {\bibinfo {volume} {5}},\ \bibinfo {pages} {L022007} (\bibinfo {year} {2023})}\BibitemShut {NoStop}%
\bibitem [{\citenamefont {Ono}\ and\ \citenamefont {Akagi}(2023)}]{ono2023photocontrol}%
  \BibitemOpen
  \bibfield  {author} {\bibinfo {author} {\bibfnamefont {A.}~\bibnamefont {Ono}}\ and\ \bibinfo {author} {\bibfnamefont {Y.}~\bibnamefont {Akagi}},\ }\bibfield  {title} {\bibinfo {title} {Photocontrol of spin scalar chirality in centrosymmetric itinerant magnets},\ }\href@noop {} {\bibfield  {journal} {\bibinfo  {journal} {Physical Review B}\ }\textbf {\bibinfo {volume} {108}},\ \bibinfo {pages} {L100407} (\bibinfo {year} {2023})}\BibitemShut {NoStop}%
\bibitem [{\citenamefont {Joseph}\ \emph {et~al.}(2024)\citenamefont {Joseph}, \citenamefont {Nair}, \citenamefont {Smith}, \citenamefont {Holland}, \citenamefont {McLellan}, \citenamefont {Boventer}, \citenamefont {Wolz}, \citenamefont {Bozhko}, \citenamefont {Flebus}, \citenamefont {Weides} \emph {et~al.}}]{joseph2024role}%
  \BibitemOpen
  \bibfield  {author} {\bibinfo {author} {\bibfnamefont {A.}~\bibnamefont {Joseph}}, \bibinfo {author} {\bibfnamefont {J.~M.}\ \bibnamefont {Nair}}, \bibinfo {author} {\bibfnamefont {M.~A.}\ \bibnamefont {Smith}}, \bibinfo {author} {\bibfnamefont {R.}~\bibnamefont {Holland}}, \bibinfo {author} {\bibfnamefont {L.~J.}\ \bibnamefont {McLellan}}, \bibinfo {author} {\bibfnamefont {I.}~\bibnamefont {Boventer}}, \bibinfo {author} {\bibfnamefont {T.}~\bibnamefont {Wolz}}, \bibinfo {author} {\bibfnamefont {D.~A.}\ \bibnamefont {Bozhko}}, \bibinfo {author} {\bibfnamefont {B.}~\bibnamefont {Flebus}}, \bibinfo {author} {\bibfnamefont {M.~P.}\ \bibnamefont {Weides}}, \emph {et~al.},\ }\bibfield  {title} {\bibinfo {title} {The role of excitation vector fields and all-polarisation state control in cavity magnonics},\ }\href@noop {} {\bibfield  {journal} {\bibinfo  {journal} {npj Spintronics}\ }\textbf {\bibinfo {volume} {2}},\ \bibinfo {pages} {1} (\bibinfo {year} {2024})}\BibitemShut {NoStop}%
\bibitem [{\citenamefont {Mochizuki}(2012)}]{mochizuki2012spin}%
  \BibitemOpen
  \bibfield  {author} {\bibinfo {author} {\bibfnamefont {M.}~\bibnamefont {Mochizuki}},\ }\bibfield  {title} {\bibinfo {title} {Spin-wave modes and their intense excitation effects in skyrmion crystals},\ }\href@noop {} {\bibfield  {journal} {\bibinfo  {journal} {Physical Review Letters}\ }\textbf {\bibinfo {volume} {108}},\ \bibinfo {pages} {017601} (\bibinfo {year} {2012})}\BibitemShut {NoStop}%
\bibitem [{\citenamefont {Finazzi}\ \emph {et~al.}(2013)\citenamefont {Finazzi}, \citenamefont {Savoini}, \citenamefont {Khorsand}, \citenamefont {Tsukamoto}, \citenamefont {Itoh}, \citenamefont {Duo}, \citenamefont {Kirilyuk}, \citenamefont {Rasing},\ and\ \citenamefont {Ezawa}}]{finazzi2013laser}%
  \BibitemOpen
  \bibfield  {author} {\bibinfo {author} {\bibfnamefont {M.}~\bibnamefont {Finazzi}}, \bibinfo {author} {\bibfnamefont {M.}~\bibnamefont {Savoini}}, \bibinfo {author} {\bibfnamefont {A.}~\bibnamefont {Khorsand}}, \bibinfo {author} {\bibfnamefont {A.}~\bibnamefont {Tsukamoto}}, \bibinfo {author} {\bibfnamefont {A.}~\bibnamefont {Itoh}}, \bibinfo {author} {\bibfnamefont {L.}~\bibnamefont {Duo}}, \bibinfo {author} {\bibfnamefont {A.}~\bibnamefont {Kirilyuk}}, \bibinfo {author} {\bibfnamefont {T.}~\bibnamefont {Rasing}},\ and\ \bibinfo {author} {\bibfnamefont {M.}~\bibnamefont {Ezawa}},\ }\bibfield  {title} {\bibinfo {title} {Laser-induced magnetic nanostructures with tunable topological properties},\ }\href@noop {} {\bibfield  {journal} {\bibinfo  {journal} {Physical Review Letters}\ }\textbf {\bibinfo {volume} {110}},\ \bibinfo {pages} {177205} (\bibinfo {year} {2013})}\BibitemShut {NoStop}%
\bibitem [{\citenamefont {Osada}\ \emph {et~al.}(2018)\citenamefont {Osada}, \citenamefont {Gloppe}, \citenamefont {Hisatomi}, \citenamefont {Noguchi}, \citenamefont {Yamazaki}, \citenamefont {Nomura}, \citenamefont {Nakamura},\ and\ \citenamefont {Usami}}]{osada2018brillouin}%
  \BibitemOpen
  \bibfield  {author} {\bibinfo {author} {\bibfnamefont {A.}~\bibnamefont {Osada}}, \bibinfo {author} {\bibfnamefont {A.}~\bibnamefont {Gloppe}}, \bibinfo {author} {\bibfnamefont {R.}~\bibnamefont {Hisatomi}}, \bibinfo {author} {\bibfnamefont {A.}~\bibnamefont {Noguchi}}, \bibinfo {author} {\bibfnamefont {R.}~\bibnamefont {Yamazaki}}, \bibinfo {author} {\bibfnamefont {M.}~\bibnamefont {Nomura}}, \bibinfo {author} {\bibfnamefont {Y.}~\bibnamefont {Nakamura}},\ and\ \bibinfo {author} {\bibfnamefont {K.}~\bibnamefont {Usami}},\ }\bibfield  {title} {\bibinfo {title} {Brillouin light scattering by magnetic quasivortices in cavity optomagnonics},\ }\href@noop {} {\bibfield  {journal} {\bibinfo  {journal} {Physical Review Letters}\ }\textbf {\bibinfo {volume} {120}},\ \bibinfo {pages} {133602} (\bibinfo {year} {2018})}\BibitemShut {NoStop}%
\bibitem [{\citenamefont {Fanciulli}\ \emph {et~al.}(2021)\citenamefont {Fanciulli}, \citenamefont {Bresteau}, \citenamefont {Vimal}, \citenamefont {Luttmann}, \citenamefont {Sacchi},\ and\ \citenamefont {Ruchon}}]{fanciulli2021electromagnetic}%
  \BibitemOpen
  \bibfield  {author} {\bibinfo {author} {\bibfnamefont {M.}~\bibnamefont {Fanciulli}}, \bibinfo {author} {\bibfnamefont {D.}~\bibnamefont {Bresteau}}, \bibinfo {author} {\bibfnamefont {M.}~\bibnamefont {Vimal}}, \bibinfo {author} {\bibfnamefont {M.}~\bibnamefont {Luttmann}}, \bibinfo {author} {\bibfnamefont {M.}~\bibnamefont {Sacchi}},\ and\ \bibinfo {author} {\bibfnamefont {T.}~\bibnamefont {Ruchon}},\ }\bibfield  {title} {\bibinfo {title} {Electromagnetic theory of helicoidal dichroism in reflection from magnetic structures},\ }\href@noop {} {\bibfield  {journal} {\bibinfo  {journal} {Physical Review A}\ }\textbf {\bibinfo {volume} {103}},\ \bibinfo {pages} {013501} (\bibinfo {year} {2021})}\BibitemShut {NoStop}%
\bibitem [{\citenamefont {Goto}\ \emph {et~al.}(2021)\citenamefont {Goto}, \citenamefont {Ishihara},\ and\ \citenamefont {Yokoshi}}]{goto2021twisted}%
  \BibitemOpen
  \bibfield  {author} {\bibinfo {author} {\bibfnamefont {Y.}~\bibnamefont {Goto}}, \bibinfo {author} {\bibfnamefont {H.}~\bibnamefont {Ishihara}},\ and\ \bibinfo {author} {\bibfnamefont {N.}~\bibnamefont {Yokoshi}},\ }\bibfield  {title} {\bibinfo {title} {Twisted light-induced spin--spin interaction in a chiral helimagnet},\ }\href@noop {} {\bibfield  {journal} {\bibinfo  {journal} {New Journal of Physics}\ }\textbf {\bibinfo {volume} {23}},\ \bibinfo {pages} {053004} (\bibinfo {year} {2021})}\BibitemShut {NoStop}%
\bibitem [{\citenamefont {W{\"a}tzel}\ \emph {et~al.}(2022)\citenamefont {W{\"a}tzel}, \citenamefont {Rebernik~Ribi{\v{c}}}, \citenamefont {Coreno}, \citenamefont {Danailov}, \citenamefont {David}, \citenamefont {Demidovich}, \citenamefont {Di~Fraia}, \citenamefont {Giannessi}, \citenamefont {Hansen}, \citenamefont {Kru{\v{s}}i{\v{c}}} \emph {et~al.}}]{watzel2022light}%
  \BibitemOpen
  \bibfield  {author} {\bibinfo {author} {\bibfnamefont {J.}~\bibnamefont {W{\"a}tzel}}, \bibinfo {author} {\bibfnamefont {P.}~\bibnamefont {Rebernik~Ribi{\v{c}}}}, \bibinfo {author} {\bibfnamefont {M.}~\bibnamefont {Coreno}}, \bibinfo {author} {\bibfnamefont {M.~B.}\ \bibnamefont {Danailov}}, \bibinfo {author} {\bibfnamefont {C.}~\bibnamefont {David}}, \bibinfo {author} {\bibfnamefont {A.}~\bibnamefont {Demidovich}}, \bibinfo {author} {\bibfnamefont {M.}~\bibnamefont {Di~Fraia}}, \bibinfo {author} {\bibfnamefont {L.}~\bibnamefont {Giannessi}}, \bibinfo {author} {\bibfnamefont {K.}~\bibnamefont {Hansen}}, \bibinfo {author} {\bibfnamefont {{\v{S}}.}~\bibnamefont {Kru{\v{s}}i{\v{c}}}}, \emph {et~al.},\ }\bibfield  {title} {\bibinfo {title} {Light-induced magnetization at the nanoscale},\ }\href@noop {} {\bibfield  {journal} {\bibinfo  {journal} {Physical Review Letters}\ }\textbf {\bibinfo {volume} {128}},\ \bibinfo {pages} {157205} (\bibinfo {year} {2022})}\BibitemShut {NoStop}%
\bibitem [{\citenamefont {Fanciulli}\ \emph {et~al.}(2022)\citenamefont {Fanciulli}, \citenamefont {Pancaldi}, \citenamefont {Pedersoli}, \citenamefont {Vimal}, \citenamefont {Bresteau}, \citenamefont {Luttmann}, \citenamefont {De~Angelis}, \citenamefont {Ribi{\v{c}}}, \citenamefont {R{\"o}sner}, \citenamefont {David} \emph {et~al.}}]{fanciulli2022observation}%
  \BibitemOpen
  \bibfield  {author} {\bibinfo {author} {\bibfnamefont {M.}~\bibnamefont {Fanciulli}}, \bibinfo {author} {\bibfnamefont {M.}~\bibnamefont {Pancaldi}}, \bibinfo {author} {\bibfnamefont {E.}~\bibnamefont {Pedersoli}}, \bibinfo {author} {\bibfnamefont {M.}~\bibnamefont {Vimal}}, \bibinfo {author} {\bibfnamefont {D.}~\bibnamefont {Bresteau}}, \bibinfo {author} {\bibfnamefont {M.}~\bibnamefont {Luttmann}}, \bibinfo {author} {\bibfnamefont {D.}~\bibnamefont {De~Angelis}}, \bibinfo {author} {\bibfnamefont {P.~R.}\ \bibnamefont {Ribi{\v{c}}}}, \bibinfo {author} {\bibfnamefont {B.}~\bibnamefont {R{\"o}sner}}, \bibinfo {author} {\bibfnamefont {C.}~\bibnamefont {David}}, \emph {et~al.},\ }\bibfield  {title} {\bibinfo {title} {Observation of magnetic helicoidal dichroism with extreme ultraviolet light vortices},\ }\href@noop {} {\bibfield  {journal} {\bibinfo  {journal} {Physical Review Letters}\ }\textbf {\bibinfo {volume} {128}},\ \bibinfo {pages} {077401} (\bibinfo {year} {2022})}\BibitemShut {NoStop}%
\bibitem [{\citenamefont {Gao}\ \emph {et~al.}(2023)\citenamefont {Gao}, \citenamefont {Prokhorenko}, \citenamefont {Nahas},\ and\ \citenamefont {Bellaiche}}]{gao2023dynamical}%
  \BibitemOpen
  \bibfield  {author} {\bibinfo {author} {\bibfnamefont {L.}~\bibnamefont {Gao}}, \bibinfo {author} {\bibfnamefont {S.}~\bibnamefont {Prokhorenko}}, \bibinfo {author} {\bibfnamefont {Y.}~\bibnamefont {Nahas}},\ and\ \bibinfo {author} {\bibfnamefont {L.}~\bibnamefont {Bellaiche}},\ }\bibfield  {title} {\bibinfo {title} {Dynamical multiferroicity and magnetic topological structures induced by the orbital angular momentum of light in a nonmagnetic material},\ }\href@noop {} {\bibfield  {journal} {\bibinfo  {journal} {Physical Review Letters}\ }\textbf {\bibinfo {volume} {131}},\ \bibinfo {pages} {196801} (\bibinfo {year} {2023})}\BibitemShut {NoStop}%
\bibitem [{\citenamefont {Rameshti}\ \emph {et~al.}(2024)\citenamefont {Rameshti}, \citenamefont {Tabatabaei},\ and\ \citenamefont {Duine}}]{rameshti2024twisted}%
  \BibitemOpen
  \bibfield  {author} {\bibinfo {author} {\bibfnamefont {B.~Z.}\ \bibnamefont {Rameshti}}, \bibinfo {author} {\bibfnamefont {S.~M.}\ \bibnamefont {Tabatabaei}},\ and\ \bibinfo {author} {\bibfnamefont {R.~A.}\ \bibnamefont {Duine}},\ }\bibfield  {title} {\bibinfo {title} {Twisted dynamics in magnetic insulators with structured light},\ }\href@noop {} {\bibfield  {journal} {\bibinfo  {journal} {Physical Review B}\ }\textbf {\bibinfo {volume} {110}},\ \bibinfo {pages} {094437} (\bibinfo {year} {2024})}\BibitemShut {NoStop}%
\bibitem [{\citenamefont {Du}\ \emph {et~al.}(2019)\citenamefont {Du}, \citenamefont {Yang}, \citenamefont {Zayats},\ and\ \citenamefont {Yuan}}]{du2019deep}%
  \BibitemOpen
  \bibfield  {author} {\bibinfo {author} {\bibfnamefont {L.}~\bibnamefont {Du}}, \bibinfo {author} {\bibfnamefont {A.}~\bibnamefont {Yang}}, \bibinfo {author} {\bibfnamefont {A.~V.}\ \bibnamefont {Zayats}},\ and\ \bibinfo {author} {\bibfnamefont {X.}~\bibnamefont {Yuan}},\ }\bibfield  {title} {\bibinfo {title} {Deep-subwavelength features of photonic skyrmions in a confined electromagnetic field with orbital angular momentum},\ }\href@noop {} {\bibfield  {journal} {\bibinfo  {journal} {Nature Physics}\ }\textbf {\bibinfo {volume} {15}},\ \bibinfo {pages} {650} (\bibinfo {year} {2019})}\BibitemShut {NoStop}%
\bibitem [{\citenamefont {S{\'a}nchez-Tejerina}\ \emph {et~al.}(2023)\citenamefont {S{\'a}nchez-Tejerina}, \citenamefont {Mart{\'\i}n-Hern{\'a}ndez}, \citenamefont {Yanes}, \citenamefont {Plaja}, \citenamefont {L{\'o}pez-D{\'\i}az},\ and\ \citenamefont {Hern{\'a}ndez-Garc{\'\i}a}}]{sanchez2023all}%
  \BibitemOpen
  \bibfield  {author} {\bibinfo {author} {\bibfnamefont {L.}~\bibnamefont {S{\'a}nchez-Tejerina}}, \bibinfo {author} {\bibfnamefont {R.}~\bibnamefont {Mart{\'\i}n-Hern{\'a}ndez}}, \bibinfo {author} {\bibfnamefont {R.}~\bibnamefont {Yanes}}, \bibinfo {author} {\bibfnamefont {L.}~\bibnamefont {Plaja}}, \bibinfo {author} {\bibfnamefont {L.}~\bibnamefont {L{\'o}pez-D{\'\i}az}},\ and\ \bibinfo {author} {\bibfnamefont {C.}~\bibnamefont {Hern{\'a}ndez-Garc{\'\i}a}},\ }\bibfield  {title} {\bibinfo {title} {All-optical nonlinear chiral ultrafast magnetization dynamics driven by circularly polarized magnetic fields},\ }\href@noop {} {\bibfield  {journal} {\bibinfo  {journal} {High Power Laser Science and Engineering}\ }\textbf {\bibinfo {volume} {11}},\ \bibinfo {pages} {e82} (\bibinfo {year} {2023})}\BibitemShut {NoStop}%
\bibitem [{\citenamefont {Roessler}\ \emph {et~al.}(2006)\citenamefont {Roessler}, \citenamefont {Bogdanov},\ and\ \citenamefont {Pfleiderer}}]{roessler2006spontaneous}%
  \BibitemOpen
  \bibfield  {author} {\bibinfo {author} {\bibfnamefont {U.~K.}\ \bibnamefont {Roessler}}, \bibinfo {author} {\bibfnamefont {A.}~\bibnamefont {Bogdanov}},\ and\ \bibinfo {author} {\bibfnamefont {C.}~\bibnamefont {Pfleiderer}},\ }\bibfield  {title} {\bibinfo {title} {Spontaneous skyrmion ground states in magnetic metals},\ }\href@noop {} {\bibfield  {journal} {\bibinfo  {journal} {Nature}\ }\textbf {\bibinfo {volume} {442}},\ \bibinfo {pages} {797} (\bibinfo {year} {2006})}\BibitemShut {NoStop}%
\bibitem [{\citenamefont {Mu{\"u}hlbauer}\ \emph {et~al.}(2009)\citenamefont {Mu{\"u}hlbauer}, \citenamefont {Binz}, \citenamefont {Jonietz}, \citenamefont {Pfleiderer}, \citenamefont {Rosch}, \citenamefont {Neubauer}, \citenamefont {Georgii},\ and\ \citenamefont {B{\"o}ni}}]{mühlbauer2009skyrmion}%
  \BibitemOpen
  \bibfield  {author} {\bibinfo {author} {\bibfnamefont {S.}~\bibnamefont {Mu{\"u}hlbauer}}, \bibinfo {author} {\bibfnamefont {B.}~\bibnamefont {Binz}}, \bibinfo {author} {\bibfnamefont {F.}~\bibnamefont {Jonietz}}, \bibinfo {author} {\bibfnamefont {C.}~\bibnamefont {Pfleiderer}}, \bibinfo {author} {\bibfnamefont {A.}~\bibnamefont {Rosch}}, \bibinfo {author} {\bibfnamefont {A.}~\bibnamefont {Neubauer}}, \bibinfo {author} {\bibfnamefont {R.}~\bibnamefont {Georgii}},\ and\ \bibinfo {author} {\bibfnamefont {P.}~\bibnamefont {B{\"o}ni}},\ }\bibfield  {title} {\bibinfo {title} {Skyrmion lattice in a chiral magnet},\ }\href@noop {} {\bibfield  {journal} {\bibinfo  {journal} {Science}\ }\textbf {\bibinfo {volume} {323}},\ \bibinfo {pages} {915} (\bibinfo {year} {2009})}\BibitemShut {NoStop}%
\bibitem [{\citenamefont {Nagaosa}\ and\ \citenamefont {Tokura}(2013)}]{nagaosa2013topological}%
  \BibitemOpen
  \bibfield  {author} {\bibinfo {author} {\bibfnamefont {N.}~\bibnamefont {Nagaosa}}\ and\ \bibinfo {author} {\bibfnamefont {Y.}~\bibnamefont {Tokura}},\ }\bibfield  {title} {\bibinfo {title} {Topological properties and dynamics of magnetic skyrmions},\ }\href@noop {} {\bibfield  {journal} {\bibinfo  {journal} {Nature Nanotechnology}\ }\textbf {\bibinfo {volume} {8}},\ \bibinfo {pages} {899} (\bibinfo {year} {2013})}\BibitemShut {NoStop}%
\bibitem [{\citenamefont {Mochizuki}\ and\ \citenamefont {Watanabe}(2015)}]{mochizuki2015writing}%
  \BibitemOpen
  \bibfield  {author} {\bibinfo {author} {\bibfnamefont {M.}~\bibnamefont {Mochizuki}}\ and\ \bibinfo {author} {\bibfnamefont {Y.}~\bibnamefont {Watanabe}},\ }\bibfield  {title} {\bibinfo {title} {Writing a skyrmion on multiferroic materials},\ }\href@noop {} {\bibfield  {journal} {\bibinfo  {journal} {Applied Physics Letters}\ }\textbf {\bibinfo {volume} {107}} (\bibinfo {year} {2015})}\BibitemShut {NoStop}%
\bibitem [{\citenamefont {Yuan}\ and\ \citenamefont {Wang}(2016)}]{yuan2016skyrmion}%
  \BibitemOpen
  \bibfield  {author} {\bibinfo {author} {\bibfnamefont {H.}~\bibnamefont {Yuan}}\ and\ \bibinfo {author} {\bibfnamefont {X.}~\bibnamefont {Wang}},\ }\bibfield  {title} {\bibinfo {title} {Skyrmion creation and manipulation by nano-second current pulses},\ }\href@noop {} {\bibfield  {journal} {\bibinfo  {journal} {Scientific Reports}\ }\textbf {\bibinfo {volume} {6}},\ \bibinfo {pages} {22638} (\bibinfo {year} {2016})}\BibitemShut {NoStop}%
\bibitem [{\citenamefont {Liu}\ \emph {et~al.}(2015)\citenamefont {Liu}, \citenamefont {Yin}, \citenamefont {Zang}, \citenamefont {Shi},\ and\ \citenamefont {Lake}}]{liu2015skyrmion}%
  \BibitemOpen
  \bibfield  {author} {\bibinfo {author} {\bibfnamefont {Y.}~\bibnamefont {Liu}}, \bibinfo {author} {\bibfnamefont {G.}~\bibnamefont {Yin}}, \bibinfo {author} {\bibfnamefont {J.}~\bibnamefont {Zang}}, \bibinfo {author} {\bibfnamefont {J.}~\bibnamefont {Shi}},\ and\ \bibinfo {author} {\bibfnamefont {R.~K.}\ \bibnamefont {Lake}},\ }\bibfield  {title} {\bibinfo {title} {Skyrmion creation and annihilation by spin waves},\ }\href@noop {} {\bibfield  {journal} {\bibinfo  {journal} {Applied Physics Letters}\ }\textbf {\bibinfo {volume} {107}} (\bibinfo {year} {2015})}\BibitemShut {NoStop}%
\bibitem [{\citenamefont {Yokouchi}\ \emph {et~al.}(2020)\citenamefont {Yokouchi}, \citenamefont {Sugimoto}, \citenamefont {Rana}, \citenamefont {Seki}, \citenamefont {Ogawa}, \citenamefont {Kasai},\ and\ \citenamefont {Otani}}]{yokouchi2020creation}%
  \BibitemOpen
  \bibfield  {author} {\bibinfo {author} {\bibfnamefont {T.}~\bibnamefont {Yokouchi}}, \bibinfo {author} {\bibfnamefont {S.}~\bibnamefont {Sugimoto}}, \bibinfo {author} {\bibfnamefont {B.}~\bibnamefont {Rana}}, \bibinfo {author} {\bibfnamefont {S.}~\bibnamefont {Seki}}, \bibinfo {author} {\bibfnamefont {N.}~\bibnamefont {Ogawa}}, \bibinfo {author} {\bibfnamefont {S.}~\bibnamefont {Kasai}},\ and\ \bibinfo {author} {\bibfnamefont {Y.}~\bibnamefont {Otani}},\ }\bibfield  {title} {\bibinfo {title} {Creation of magnetic skyrmions by surface acoustic waves},\ }\href@noop {} {\bibfield  {journal} {\bibinfo  {journal} {Nature Nanotechnology}\ }\textbf {\bibinfo {volume} {15}},\ \bibinfo {pages} {361} (\bibinfo {year} {2020})}\BibitemShut {NoStop}%
\bibitem [{\citenamefont {Ogawa}\ \emph {et~al.}(2015)\citenamefont {Ogawa}, \citenamefont {Seki},\ and\ \citenamefont {Tokura}}]{ogawa2015ultrafast}%
  \BibitemOpen
  \bibfield  {author} {\bibinfo {author} {\bibfnamefont {N.}~\bibnamefont {Ogawa}}, \bibinfo {author} {\bibfnamefont {S.}~\bibnamefont {Seki}},\ and\ \bibinfo {author} {\bibfnamefont {Y.}~\bibnamefont {Tokura}},\ }\bibfield  {title} {\bibinfo {title} {Ultrafast optical excitation of magnetic skyrmions},\ }\href@noop {} {\bibfield  {journal} {\bibinfo  {journal} {Scientific Reports}\ }\textbf {\bibinfo {volume} {5}},\ \bibinfo {pages} {9552} (\bibinfo {year} {2015})}\BibitemShut {NoStop}%
\bibitem [{\citenamefont {B{\"u}ttner}\ \emph {et~al.}(2021)\citenamefont {B{\"u}ttner}, \citenamefont {Pfau}, \citenamefont {B{\"o}ttcher}, \citenamefont {Schneider}, \citenamefont {Mercurio}, \citenamefont {G{\"u}nther}, \citenamefont {Hessing}, \citenamefont {Klose}, \citenamefont {Wittmann}, \citenamefont {Gerlinger} \emph {et~al.}}]{buttner2021observation}%
  \BibitemOpen
  \bibfield  {author} {\bibinfo {author} {\bibfnamefont {F.}~\bibnamefont {B{\"u}ttner}}, \bibinfo {author} {\bibfnamefont {B.}~\bibnamefont {Pfau}}, \bibinfo {author} {\bibfnamefont {M.}~\bibnamefont {B{\"o}ttcher}}, \bibinfo {author} {\bibfnamefont {M.}~\bibnamefont {Schneider}}, \bibinfo {author} {\bibfnamefont {G.}~\bibnamefont {Mercurio}}, \bibinfo {author} {\bibfnamefont {C.~M.}\ \bibnamefont {G{\"u}nther}}, \bibinfo {author} {\bibfnamefont {P.}~\bibnamefont {Hessing}}, \bibinfo {author} {\bibfnamefont {C.}~\bibnamefont {Klose}}, \bibinfo {author} {\bibfnamefont {A.}~\bibnamefont {Wittmann}}, \bibinfo {author} {\bibfnamefont {K.}~\bibnamefont {Gerlinger}}, \emph {et~al.},\ }\bibfield  {title} {\bibinfo {title} {Observation of fluctuation-mediated picosecond nucleation of a topological phase},\ }\href@noop {} {\bibfield  {journal} {\bibinfo  {journal} {Nature Materials}\ }\textbf {\bibinfo {volume} {20}},\ \bibinfo {pages} {30} (\bibinfo {year} {2021})}\BibitemShut {NoStop}%
\bibitem [{\citenamefont {Koshibae}\ and\ \citenamefont {Nagaosa}(2014)}]{koshibae2014creation}%
  \BibitemOpen
  \bibfield  {author} {\bibinfo {author} {\bibfnamefont {W.}~\bibnamefont {Koshibae}}\ and\ \bibinfo {author} {\bibfnamefont {N.}~\bibnamefont {Nagaosa}},\ }\bibfield  {title} {\bibinfo {title} {Creation of skyrmions and antiskyrmions by local heating},\ }\href@noop {} {\bibfield  {journal} {\bibinfo  {journal} {Nature Communications}\ }\textbf {\bibinfo {volume} {5}},\ \bibinfo {pages} {5148} (\bibinfo {year} {2014})}\BibitemShut {NoStop}%
\bibitem [{\citenamefont {Liu}\ and\ \citenamefont {Li}(2013)}]{liu2013mechanism}%
  \BibitemOpen
  \bibfield  {author} {\bibinfo {author} {\bibfnamefont {Y.-H.}\ \bibnamefont {Liu}}\ and\ \bibinfo {author} {\bibfnamefont {Y.-Q.}\ \bibnamefont {Li}},\ }\bibfield  {title} {\bibinfo {title} {A mechanism to pin skyrmions in chiral magnets},\ }\href@noop {} {\bibfield  {journal} {\bibinfo  {journal} {Journal of Physics: Condensed Matter}\ }\textbf {\bibinfo {volume} {25}},\ \bibinfo {pages} {076005} (\bibinfo {year} {2013})}\BibitemShut {NoStop}%
\bibitem [{\citenamefont {Guan}\ \emph {et~al.}(2022)\citenamefont {Guan}, \citenamefont {Yang}, \citenamefont {Jin}, \citenamefont {Liu}, \citenamefont {Liu}, \citenamefont {Hou}, \citenamefont {Chen}, \citenamefont {Fan}, \citenamefont {Zeng}, \citenamefont {Lu} \emph {et~al.}}]{guan2022unidirectional}%
  \BibitemOpen
  \bibfield  {author} {\bibinfo {author} {\bibfnamefont {S.}~\bibnamefont {Guan}}, \bibinfo {author} {\bibfnamefont {Y.}~\bibnamefont {Yang}}, \bibinfo {author} {\bibfnamefont {Z.}~\bibnamefont {Jin}}, \bibinfo {author} {\bibfnamefont {T.}~\bibnamefont {Liu}}, \bibinfo {author} {\bibfnamefont {Y.}~\bibnamefont {Liu}}, \bibinfo {author} {\bibfnamefont {Z.}~\bibnamefont {Hou}}, \bibinfo {author} {\bibfnamefont {D.}~\bibnamefont {Chen}}, \bibinfo {author} {\bibfnamefont {Z.}~\bibnamefont {Fan}}, \bibinfo {author} {\bibfnamefont {M.}~\bibnamefont {Zeng}}, \bibinfo {author} {\bibfnamefont {X.}~\bibnamefont {Lu}}, \emph {et~al.},\ }\bibfield  {title} {\bibinfo {title} {Unidirectional localization and track-selection of antiferromagnetic skyrmions through tuning magnetocrystalline anisotropy barriers},\ }\href@noop {} {\bibfield  {journal} {\bibinfo  {journal} {Journal of Magnetism and Magnetic Materials}\ }\textbf {\bibinfo {volume} {546}},\ \bibinfo {pages} {168852} (\bibinfo {year} {2022})}\BibitemShut {NoStop}%
\bibitem [{\citenamefont {Guan}\ \emph {et~al.}(2021)\citenamefont {Guan}, \citenamefont {Yang}, \citenamefont {Jin}, \citenamefont {Liu}, \citenamefont {Liu},\ and\ \citenamefont {Qin}}]{guan2021suppression}%
  \BibitemOpen
  \bibfield  {author} {\bibinfo {author} {\bibfnamefont {S.}~\bibnamefont {Guan}}, \bibinfo {author} {\bibfnamefont {Y.}~\bibnamefont {Yang}}, \bibinfo {author} {\bibfnamefont {Z.}~\bibnamefont {Jin}}, \bibinfo {author} {\bibfnamefont {T.}~\bibnamefont {Liu}}, \bibinfo {author} {\bibfnamefont {Y.}~\bibnamefont {Liu}},\ and\ \bibinfo {author} {\bibfnamefont {M.}~\bibnamefont {Qin}},\ }\bibfield  {title} {\bibinfo {title} {Suppression of skyrmion hall motion in antiferromagnets driven by circularly polarized spin waves},\ }\href@noop {} {\bibfield  {journal} {\bibinfo  {journal} {Frontiers in Physics}\ }\textbf {\bibinfo {volume} {9}},\ \bibinfo {pages} {754869} (\bibinfo {year} {2021})}\BibitemShut {NoStop}%
\bibitem [{\citenamefont {Li}\ \emph {et~al.}(2018)\citenamefont {Li}, \citenamefont {Xia}, \citenamefont {Zhang}, \citenamefont {Ezawa}, \citenamefont {Kang}, \citenamefont {Liu}, \citenamefont {Zhou},\ and\ \citenamefont {Zhao}}]{li2018dynamics}%
  \BibitemOpen
  \bibfield  {author} {\bibinfo {author} {\bibfnamefont {S.}~\bibnamefont {Li}}, \bibinfo {author} {\bibfnamefont {J.}~\bibnamefont {Xia}}, \bibinfo {author} {\bibfnamefont {X.}~\bibnamefont {Zhang}}, \bibinfo {author} {\bibfnamefont {M.}~\bibnamefont {Ezawa}}, \bibinfo {author} {\bibfnamefont {W.}~\bibnamefont {Kang}}, \bibinfo {author} {\bibfnamefont {X.}~\bibnamefont {Liu}}, \bibinfo {author} {\bibfnamefont {Y.}~\bibnamefont {Zhou}},\ and\ \bibinfo {author} {\bibfnamefont {W.}~\bibnamefont {Zhao}},\ }\bibfield  {title} {\bibinfo {title} {Dynamics of a magnetic skyrmionium driven by spin waves},\ }\href@noop {} {\bibfield  {journal} {\bibinfo  {journal} {Applied Physics Letters}\ }\textbf {\bibinfo {volume} {112}} (\bibinfo {year} {2018})}\BibitemShut {NoStop}%
\bibitem [{\citenamefont {Zhang}\ \emph {et~al.}(2015)\citenamefont {Zhang}, \citenamefont {Ezawa}, \citenamefont {Xiao}, \citenamefont {Zhao}, \citenamefont {Liu},\ and\ \citenamefont {Zhou}}]{zhang2015all}%
  \BibitemOpen
  \bibfield  {author} {\bibinfo {author} {\bibfnamefont {X.}~\bibnamefont {Zhang}}, \bibinfo {author} {\bibfnamefont {M.}~\bibnamefont {Ezawa}}, \bibinfo {author} {\bibfnamefont {D.}~\bibnamefont {Xiao}}, \bibinfo {author} {\bibfnamefont {G.}~\bibnamefont {Zhao}}, \bibinfo {author} {\bibfnamefont {Y.}~\bibnamefont {Liu}},\ and\ \bibinfo {author} {\bibfnamefont {Y.}~\bibnamefont {Zhou}},\ }\bibfield  {title} {\bibinfo {title} {All-magnetic control of skyrmions in nanowires by a spin wave},\ }\href@noop {} {\bibfield  {journal} {\bibinfo  {journal} {Nanotechnology}\ }\textbf {\bibinfo {volume} {26}},\ \bibinfo {pages} {225701} (\bibinfo {year} {2015})}\BibitemShut {NoStop}%
\bibitem [{\citenamefont {Kong}\ and\ \citenamefont {Zang}(2013)}]{kong2013dynamics}%
  \BibitemOpen
  \bibfield  {author} {\bibinfo {author} {\bibfnamefont {L.}~\bibnamefont {Kong}}\ and\ \bibinfo {author} {\bibfnamefont {J.}~\bibnamefont {Zang}},\ }\bibfield  {title} {\bibinfo {title} {Dynamics of an insulating skyrmion under a temperature gradient},\ }\href@noop {} {\bibfield  {journal} {\bibinfo  {journal} {Physical Review Letters}\ }\textbf {\bibinfo {volume} {111}},\ \bibinfo {pages} {067203} (\bibinfo {year} {2013})}\BibitemShut {NoStop}%
\bibitem [{\citenamefont {Kong}\ \emph {et~al.}(2021)\citenamefont {Kong}, \citenamefont {Chen}, \citenamefont {Wang}, \citenamefont {Song},\ and\ \citenamefont {Du}}]{kong2021dynamics}%
  \BibitemOpen
  \bibfield  {author} {\bibinfo {author} {\bibfnamefont {L.}~\bibnamefont {Kong}}, \bibinfo {author} {\bibfnamefont {X.}~\bibnamefont {Chen}}, \bibinfo {author} {\bibfnamefont {W.}~\bibnamefont {Wang}}, \bibinfo {author} {\bibfnamefont {D.}~\bibnamefont {Song}},\ and\ \bibinfo {author} {\bibfnamefont {H.}~\bibnamefont {Du}},\ }\bibfield  {title} {\bibinfo {title} {Dynamics of interstitial skyrmions in the presence of temperature gradients},\ }\href@noop {} {\bibfield  {journal} {\bibinfo  {journal} {Physical Review B}\ }\textbf {\bibinfo {volume} {104}},\ \bibinfo {pages} {214407} (\bibinfo {year} {2021})}\BibitemShut {NoStop}%
\bibitem [{\citenamefont {Qin}\ \emph {et~al.}(2022)\citenamefont {Qin}, \citenamefont {Zhang}, \citenamefont {Zhang}, \citenamefont {Pei}, \citenamefont {Yang}, \citenamefont {Xu}, \citenamefont {Zhou}, \citenamefont {Wu}, \citenamefont {Du},\ and\ \citenamefont {Che}}]{qin2022dynamics}%
  \BibitemOpen
  \bibfield  {author} {\bibinfo {author} {\bibfnamefont {G.}~\bibnamefont {Qin}}, \bibinfo {author} {\bibfnamefont {X.}~\bibnamefont {Zhang}}, \bibinfo {author} {\bibfnamefont {R.}~\bibnamefont {Zhang}}, \bibinfo {author} {\bibfnamefont {K.}~\bibnamefont {Pei}}, \bibinfo {author} {\bibfnamefont {C.}~\bibnamefont {Yang}}, \bibinfo {author} {\bibfnamefont {C.}~\bibnamefont {Xu}}, \bibinfo {author} {\bibfnamefont {Y.}~\bibnamefont {Zhou}}, \bibinfo {author} {\bibfnamefont {Y.}~\bibnamefont {Wu}}, \bibinfo {author} {\bibfnamefont {H.}~\bibnamefont {Du}},\ and\ \bibinfo {author} {\bibfnamefont {R.}~\bibnamefont {Che}},\ }\bibfield  {title} {\bibinfo {title} {Dynamics of magnetic skyrmions driven by a temperature gradient in a chiral magnet fege},\ }\href@noop {} {\bibfield  {journal} {\bibinfo  {journal} {Physical Review B}\ }\textbf {\bibinfo {volume} {106}},\ \bibinfo {pages} {024415} (\bibinfo {year} {2022})}\BibitemShut {NoStop}%
\bibitem [{\citenamefont {Raimondo}\ \emph {et~al.}(2022)\citenamefont {Raimondo}, \citenamefont {Saugar}, \citenamefont {Barker}, \citenamefont {Rodrigues}, \citenamefont {Giordano}, \citenamefont {Carpentieri}, \citenamefont {Jiang}, \citenamefont {Chubykalo-Fesenko}, \citenamefont {Tomasello},\ and\ \citenamefont {Finocchio}}]{raimondo2022temperature}%
  \BibitemOpen
  \bibfield  {author} {\bibinfo {author} {\bibfnamefont {E.}~\bibnamefont {Raimondo}}, \bibinfo {author} {\bibfnamefont {E.}~\bibnamefont {Saugar}}, \bibinfo {author} {\bibfnamefont {J.}~\bibnamefont {Barker}}, \bibinfo {author} {\bibfnamefont {D.}~\bibnamefont {Rodrigues}}, \bibinfo {author} {\bibfnamefont {A.}~\bibnamefont {Giordano}}, \bibinfo {author} {\bibfnamefont {M.}~\bibnamefont {Carpentieri}}, \bibinfo {author} {\bibfnamefont {W.}~\bibnamefont {Jiang}}, \bibinfo {author} {\bibfnamefont {O.}~\bibnamefont {Chubykalo-Fesenko}}, \bibinfo {author} {\bibfnamefont {R.}~\bibnamefont {Tomasello}},\ and\ \bibinfo {author} {\bibfnamefont {G.}~\bibnamefont {Finocchio}},\ }\bibfield  {title} {\bibinfo {title} {Temperature-gradient-driven magnetic skyrmion motion},\ }\href@noop {} {\bibfield  {journal} {\bibinfo  {journal} {Physical Review Applied}\ }\textbf {\bibinfo {volume} {18}},\ \bibinfo {pages} {024062} (\bibinfo {year} {2022})}\BibitemShut {NoStop}%
\bibitem [{\citenamefont {Berruto}\ \emph {et~al.}(2018)\citenamefont {Berruto}, \citenamefont {Madan}, \citenamefont {Murooka}, \citenamefont {Vanacore}, \citenamefont {Pomarico}, \citenamefont {Rajeswari}, \citenamefont {Lamb}, \citenamefont {Huang}, \citenamefont {Kruchkov}, \citenamefont {Togawa} \emph {et~al.}}]{berruto2018laser}%
  \BibitemOpen
  \bibfield  {author} {\bibinfo {author} {\bibfnamefont {G.}~\bibnamefont {Berruto}}, \bibinfo {author} {\bibfnamefont {I.}~\bibnamefont {Madan}}, \bibinfo {author} {\bibfnamefont {Y.}~\bibnamefont {Murooka}}, \bibinfo {author} {\bibfnamefont {G.}~\bibnamefont {Vanacore}}, \bibinfo {author} {\bibfnamefont {E.}~\bibnamefont {Pomarico}}, \bibinfo {author} {\bibfnamefont {J.}~\bibnamefont {Rajeswari}}, \bibinfo {author} {\bibfnamefont {R.}~\bibnamefont {Lamb}}, \bibinfo {author} {\bibfnamefont {P.}~\bibnamefont {Huang}}, \bibinfo {author} {\bibfnamefont {A.}~\bibnamefont {Kruchkov}}, \bibinfo {author} {\bibfnamefont {Y.}~\bibnamefont {Togawa}}, \emph {et~al.},\ }\bibfield  {title} {\bibinfo {title} {Laser-induced skyrmion writing and erasing in an ultrafast cryo-lorentz transmission electron microscope},\ }\href@noop {} {\bibfield  {journal} {\bibinfo  {journal} {Physical Review Letters}\ }\textbf {\bibinfo {volume} {120}},\ \bibinfo {pages} {117201} (\bibinfo {year} {2018})}\BibitemShut {NoStop}%
\bibitem [{\citenamefont {Yudin}\ \emph {et~al.}(2017)\citenamefont {Yudin}, \citenamefont {Gulevich},\ and\ \citenamefont {Titov}}]{yudin2017light}%
  \BibitemOpen
  \bibfield  {author} {\bibinfo {author} {\bibfnamefont {D.}~\bibnamefont {Yudin}}, \bibinfo {author} {\bibfnamefont {D.~R.}\ \bibnamefont {Gulevich}},\ and\ \bibinfo {author} {\bibfnamefont {M.}~\bibnamefont {Titov}},\ }\bibfield  {title} {\bibinfo {title} {Light-induced anisotropic skyrmion and stripe phases in a rashba ferromagnet},\ }\href@noop {} {\bibfield  {journal} {\bibinfo  {journal} {Physical Review Letters}\ }\textbf {\bibinfo {volume} {119}},\ \bibinfo {pages} {147202} (\bibinfo {year} {2017})}\BibitemShut {NoStop}%
\bibitem [{\citenamefont {Flovik}\ \emph {et~al.}(2017)\citenamefont {Flovik}, \citenamefont {Qaiumzadeh}, \citenamefont {Nandy}, \citenamefont {Heo},\ and\ \citenamefont {Rasing}}]{flovik2017generation}%
  \BibitemOpen
  \bibfield  {author} {\bibinfo {author} {\bibfnamefont {V.}~\bibnamefont {Flovik}}, \bibinfo {author} {\bibfnamefont {A.}~\bibnamefont {Qaiumzadeh}}, \bibinfo {author} {\bibfnamefont {A.~K.}\ \bibnamefont {Nandy}}, \bibinfo {author} {\bibfnamefont {C.}~\bibnamefont {Heo}},\ and\ \bibinfo {author} {\bibfnamefont {T.}~\bibnamefont {Rasing}},\ }\bibfield  {title} {\bibinfo {title} {Generation of single skyrmions by picosecond magnetic field pulses},\ }\href@noop {} {\bibfield  {journal} {\bibinfo  {journal} {Physical Review B}\ }\textbf {\bibinfo {volume} {96}},\ \bibinfo {pages} {140411} (\bibinfo {year} {2017})}\BibitemShut {NoStop}%
\bibitem [{\citenamefont {Je}\ \emph {et~al.}(2018)\citenamefont {Je}, \citenamefont {Vallobra}, \citenamefont {Srivastava}, \citenamefont {Rojas-S{\'a}nchez}, \citenamefont {Pham}, \citenamefont {Hehn}, \citenamefont {Malinowski}, \citenamefont {Baraduc}, \citenamefont {Auffret}, \citenamefont {Gaudin} \emph {et~al.}}]{je2018creation}%
  \BibitemOpen
  \bibfield  {author} {\bibinfo {author} {\bibfnamefont {S.-G.}\ \bibnamefont {Je}}, \bibinfo {author} {\bibfnamefont {P.}~\bibnamefont {Vallobra}}, \bibinfo {author} {\bibfnamefont {T.}~\bibnamefont {Srivastava}}, \bibinfo {author} {\bibfnamefont {J.-C.}\ \bibnamefont {Rojas-S{\'a}nchez}}, \bibinfo {author} {\bibfnamefont {T.~H.}\ \bibnamefont {Pham}}, \bibinfo {author} {\bibfnamefont {M.}~\bibnamefont {Hehn}}, \bibinfo {author} {\bibfnamefont {G.}~\bibnamefont {Malinowski}}, \bibinfo {author} {\bibfnamefont {C.}~\bibnamefont {Baraduc}}, \bibinfo {author} {\bibfnamefont {S.}~\bibnamefont {Auffret}}, \bibinfo {author} {\bibfnamefont {G.}~\bibnamefont {Gaudin}}, \emph {et~al.},\ }\bibfield  {title} {\bibinfo {title} {Creation of magnetic skyrmion bubble lattices by ultrafast laser in ultrathin films},\ }\href@noop {} {\bibfield  {journal} {\bibinfo  {journal} {Nano Letters}\ }\textbf {\bibinfo {volume} {18}},\ \bibinfo {pages} {7362} (\bibinfo {year} {2018})}\BibitemShut {NoStop}%
\bibitem [{\citenamefont {Khoshlahni}\ \emph {et~al.}(2019)\citenamefont {Khoshlahni}, \citenamefont {Qaiumzadeh}, \citenamefont {Bergman},\ and\ \citenamefont {Brataas}}]{khoshlahni2019ultrafast}%
  \BibitemOpen
  \bibfield  {author} {\bibinfo {author} {\bibfnamefont {R.}~\bibnamefont {Khoshlahni}}, \bibinfo {author} {\bibfnamefont {A.}~\bibnamefont {Qaiumzadeh}}, \bibinfo {author} {\bibfnamefont {A.}~\bibnamefont {Bergman}},\ and\ \bibinfo {author} {\bibfnamefont {A.}~\bibnamefont {Brataas}},\ }\bibfield  {title} {\bibinfo {title} {Ultrafast generation and dynamics of isolated skyrmions in antiferromagnetic insulators},\ }\href@noop {} {\bibfield  {journal} {\bibinfo  {journal} {Physical Review B}\ }\textbf {\bibinfo {volume} {99}},\ \bibinfo {pages} {054423} (\bibinfo {year} {2019})}\BibitemShut {NoStop}%
\bibitem [{\citenamefont {Vi{\~n}as~Bostr{\"o}m}\ \emph {et~al.}(2022)\citenamefont {Vi{\~n}as~Bostr{\"o}m}, \citenamefont {Rubio},\ and\ \citenamefont {Verdozzi}}]{vinas2022microscopic}%
  \BibitemOpen
  \bibfield  {author} {\bibinfo {author} {\bibfnamefont {E.}~\bibnamefont {Vi{\~n}as~Bostr{\"o}m}}, \bibinfo {author} {\bibfnamefont {A.}~\bibnamefont {Rubio}},\ and\ \bibinfo {author} {\bibfnamefont {C.}~\bibnamefont {Verdozzi}},\ }\bibfield  {title} {\bibinfo {title} {Microscopic theory of light-induced ultrafast skyrmion excitation in transition metal films},\ }\href@noop {} {\bibfield  {journal} {\bibinfo  {journal} {npj Computational Materials}\ }\textbf {\bibinfo {volume} {8}},\ \bibinfo {pages} {62} (\bibinfo {year} {2022})}\BibitemShut {NoStop}%
\bibitem [{\citenamefont {Kern}\ \emph {et~al.}(2022)\citenamefont {Kern}, \citenamefont {Pfau}, \citenamefont {Schneider}, \citenamefont {Gerlinger}, \citenamefont {Deinhart}, \citenamefont {Wittrock}, \citenamefont {Sidiropoulos}, \citenamefont {Engel}, \citenamefont {Will}, \citenamefont {G{\"u}nther} \emph {et~al.}}]{kern2022tailoring}%
  \BibitemOpen
  \bibfield  {author} {\bibinfo {author} {\bibfnamefont {L.-M.}\ \bibnamefont {Kern}}, \bibinfo {author} {\bibfnamefont {B.}~\bibnamefont {Pfau}}, \bibinfo {author} {\bibfnamefont {M.}~\bibnamefont {Schneider}}, \bibinfo {author} {\bibfnamefont {K.}~\bibnamefont {Gerlinger}}, \bibinfo {author} {\bibfnamefont {V.}~\bibnamefont {Deinhart}}, \bibinfo {author} {\bibfnamefont {S.}~\bibnamefont {Wittrock}}, \bibinfo {author} {\bibfnamefont {T.}~\bibnamefont {Sidiropoulos}}, \bibinfo {author} {\bibfnamefont {D.}~\bibnamefont {Engel}}, \bibinfo {author} {\bibfnamefont {I.}~\bibnamefont {Will}}, \bibinfo {author} {\bibfnamefont {C.~M.}\ \bibnamefont {G{\"u}nther}}, \emph {et~al.},\ }\bibfield  {title} {\bibinfo {title} {Tailoring optical excitation to control magnetic skyrmion nucleation},\ }\href@noop {} {\bibfield  {journal} {\bibinfo  {journal} {Physical Review B}\ }\textbf {\bibinfo {volume} {106}},\ \bibinfo {pages} {054435} (\bibinfo {year} {2022})}\BibitemShut {NoStop}%
\bibitem [{\citenamefont {Truc}\ \emph {et~al.}(2023)\citenamefont {Truc}, \citenamefont {Sapozhnik}, \citenamefont {Tengdin}, \citenamefont {Vi{\~n}as~Bostr{\"o}m}, \citenamefont {Sch{\"o}nenberger}, \citenamefont {Gargiulo}, \citenamefont {Madan}, \citenamefont {LaGrange}, \citenamefont {Magrez}, \citenamefont {Verdozzi} \emph {et~al.}}]{truc2023light}%
  \BibitemOpen
  \bibfield  {author} {\bibinfo {author} {\bibfnamefont {B.}~\bibnamefont {Truc}}, \bibinfo {author} {\bibfnamefont {A.~A.}\ \bibnamefont {Sapozhnik}}, \bibinfo {author} {\bibfnamefont {P.}~\bibnamefont {Tengdin}}, \bibinfo {author} {\bibfnamefont {E.}~\bibnamefont {Vi{\~n}as~Bostr{\"o}m}}, \bibinfo {author} {\bibfnamefont {T.}~\bibnamefont {Sch{\"o}nenberger}}, \bibinfo {author} {\bibfnamefont {S.}~\bibnamefont {Gargiulo}}, \bibinfo {author} {\bibfnamefont {I.}~\bibnamefont {Madan}}, \bibinfo {author} {\bibfnamefont {T.}~\bibnamefont {LaGrange}}, \bibinfo {author} {\bibfnamefont {A.}~\bibnamefont {Magrez}}, \bibinfo {author} {\bibfnamefont {C.}~\bibnamefont {Verdozzi}}, \emph {et~al.},\ }\bibfield  {title} {\bibinfo {title} {Light-induced metastable hidden skyrmion phase in the mott insulator cu2oseo3},\ }\href@noop {} {\bibfield  {journal} {\bibinfo  {journal} {Advanced Materials}\ }\textbf {\bibinfo {volume} {35}},\ \bibinfo {pages} {2304197} (\bibinfo {year} {2023})}\BibitemShut {NoStop}%
\bibitem [{\citenamefont {Li}\ \emph {et~al.}(2024)\citenamefont {Li}, \citenamefont {Zhang}, \citenamefont {Li}, \citenamefont {Guo}, \citenamefont {Wang}, \citenamefont {Deng}, \citenamefont {Hu}, \citenamefont {Hu}, \citenamefont {Liu}, \citenamefont {Qin} \emph {et~al.}}]{li2024room}%
  \BibitemOpen
  \bibfield  {author} {\bibinfo {author} {\bibfnamefont {Z.}~\bibnamefont {Li}}, \bibinfo {author} {\bibfnamefont {H.}~\bibnamefont {Zhang}}, \bibinfo {author} {\bibfnamefont {G.}~\bibnamefont {Li}}, \bibinfo {author} {\bibfnamefont {J.}~\bibnamefont {Guo}}, \bibinfo {author} {\bibfnamefont {Q.}~\bibnamefont {Wang}}, \bibinfo {author} {\bibfnamefont {Y.}~\bibnamefont {Deng}}, \bibinfo {author} {\bibfnamefont {Y.}~\bibnamefont {Hu}}, \bibinfo {author} {\bibfnamefont {X.}~\bibnamefont {Hu}}, \bibinfo {author} {\bibfnamefont {C.}~\bibnamefont {Liu}}, \bibinfo {author} {\bibfnamefont {M.}~\bibnamefont {Qin}}, \emph {et~al.},\ }\bibfield  {title} {\bibinfo {title} {Room-temperature sub-100 nm n{\'e}el-type skyrmions in non-stoichiometric van der waals ferromagnet fe3-x gate2 with ultrafast laser writability},\ }\href@noop {} {\bibfield  {journal} {\bibinfo  {journal} {Nature Communications}\ }\textbf {\bibinfo {volume} {15}},\ \bibinfo {pages} {1017} (\bibinfo {year} {2024})}\BibitemShut {NoStop}%
\bibitem [{\citenamefont {Yambe}\ and\ \citenamefont {Hayami}(2024)}]{yambe2024dynamical}%
  \BibitemOpen
  \bibfield  {author} {\bibinfo {author} {\bibfnamefont {R.}~\bibnamefont {Yambe}}\ and\ \bibinfo {author} {\bibfnamefont {S.}~\bibnamefont {Hayami}},\ }\bibfield  {title} {\bibinfo {title} {Dynamical generation of skyrmion and bimeron crystals by a circularly polarized electric field in frustrated magnets},\ }\href@noop {} {\bibfield  {journal} {\bibinfo  {journal} {Physical Review B}\ }\textbf {\bibinfo {volume} {110}},\ \bibinfo {pages} {014428} (\bibinfo {year} {2024})}\BibitemShut {NoStop}%
\bibitem [{\citenamefont {Fujita}\ and\ \citenamefont {Sato}(2017{\natexlab{a}})}]{fujita2017ultrafast}%
  \BibitemOpen
  \bibfield  {author} {\bibinfo {author} {\bibfnamefont {H.}~\bibnamefont {Fujita}}\ and\ \bibinfo {author} {\bibfnamefont {M.}~\bibnamefont {Sato}},\ }\bibfield  {title} {\bibinfo {title} {Ultrafast generation of skyrmionic defects with vortex beams: Printing laser profiles on magnets},\ }\href@noop {} {\bibfield  {journal} {\bibinfo  {journal} {Physical Review B}\ }\textbf {\bibinfo {volume} {95}},\ \bibinfo {pages} {054421} (\bibinfo {year} {2017}{\natexlab{a}})}\BibitemShut {NoStop}%
\bibitem [{\citenamefont {Fujita}\ and\ \citenamefont {Sato}(2017{\natexlab{b}})}]{fujita2017encoding}%
  \BibitemOpen
  \bibfield  {author} {\bibinfo {author} {\bibfnamefont {H.}~\bibnamefont {Fujita}}\ and\ \bibinfo {author} {\bibfnamefont {M.}~\bibnamefont {Sato}},\ }\bibfield  {title} {\bibinfo {title} {Encoding orbital angular momentum of light in magnets},\ }\href@noop {} {\bibfield  {journal} {\bibinfo  {journal} {Physical Review B}\ }\textbf {\bibinfo {volume} {96}},\ \bibinfo {pages} {060407} (\bibinfo {year} {2017}{\natexlab{b}})}\BibitemShut {NoStop}%
\bibitem [{\citenamefont {Yang}\ \emph {et~al.}(2018)\citenamefont {Yang}, \citenamefont {Yang}, \citenamefont {Cao},\ and\ \citenamefont {Yan}}]{yang2018photonic}%
  \BibitemOpen
  \bibfield  {author} {\bibinfo {author} {\bibfnamefont {W.}~\bibnamefont {Yang}}, \bibinfo {author} {\bibfnamefont {H.}~\bibnamefont {Yang}}, \bibinfo {author} {\bibfnamefont {Y.}~\bibnamefont {Cao}},\ and\ \bibinfo {author} {\bibfnamefont {P.}~\bibnamefont {Yan}},\ }\bibfield  {title} {\bibinfo {title} {Photonic orbital angular momentum transfer and magnetic skyrmion rotation},\ }\href@noop {} {\bibfield  {journal} {\bibinfo  {journal} {Optics Express}\ }\textbf {\bibinfo {volume} {26}},\ \bibinfo {pages} {8778} (\bibinfo {year} {2018})}\BibitemShut {NoStop}%
\bibitem [{\citenamefont {Jiang}\ \emph {et~al.}(2024)\citenamefont {Jiang}, \citenamefont {Vallobra}, \citenamefont {Zhang}, \citenamefont {Xu}, \citenamefont {Zhan},\ and\ \citenamefont {Zhao}}]{jiang2024generation}%
  \BibitemOpen
  \bibfield  {author} {\bibinfo {author} {\bibfnamefont {Y.}~\bibnamefont {Jiang}}, \bibinfo {author} {\bibfnamefont {P.}~\bibnamefont {Vallobra}}, \bibinfo {author} {\bibfnamefont {X.}~\bibnamefont {Zhang}}, \bibinfo {author} {\bibfnamefont {Y.}~\bibnamefont {Xu}}, \bibinfo {author} {\bibfnamefont {Q.}~\bibnamefont {Zhan}},\ and\ \bibinfo {author} {\bibfnamefont {W.}~\bibnamefont {Zhao}},\ }\bibfield  {title} {\bibinfo {title} {Generation of topological spin textures using light-induced radially polarized magnetic field},\ }\href@noop {} {\bibfield  {journal} {\bibinfo  {journal} {Physical Review Applied}\ }\textbf {\bibinfo {volume} {21}},\ \bibinfo {pages} {044041} (\bibinfo {year} {2024})}\BibitemShut {NoStop}%
\bibitem [{\citenamefont {Assouline}\ and\ \citenamefont {Capua}(2024)}]{assouline2024helicity}%
  \BibitemOpen
  \bibfield  {author} {\bibinfo {author} {\bibfnamefont {B.}~\bibnamefont {Assouline}}\ and\ \bibinfo {author} {\bibfnamefont {A.}~\bibnamefont {Capua}},\ }\bibfield  {title} {\bibinfo {title} {Helicity-dependent optical control of the magnetization state emerging from the landau-lifshitz-gilbert equation},\ }\href@noop {} {\bibfield  {journal} {\bibinfo  {journal} {Physical Review Research}\ }\textbf {\bibinfo {volume} {6}},\ \bibinfo {pages} {013012} (\bibinfo {year} {2024})}\BibitemShut {NoStop}%
\bibitem [{\citenamefont {Zhang}\ \emph {et~al.}(2016)\citenamefont {Zhang}, \citenamefont {Xia}, \citenamefont {Zhou}, \citenamefont {Wang}, \citenamefont {Liu}, \citenamefont {Zhao},\ and\ \citenamefont {Ezawa}}]{zhang2016control}%
  \BibitemOpen
  \bibfield  {author} {\bibinfo {author} {\bibfnamefont {X.}~\bibnamefont {Zhang}}, \bibinfo {author} {\bibfnamefont {J.}~\bibnamefont {Xia}}, \bibinfo {author} {\bibfnamefont {Y.}~\bibnamefont {Zhou}}, \bibinfo {author} {\bibfnamefont {D.}~\bibnamefont {Wang}}, \bibinfo {author} {\bibfnamefont {X.}~\bibnamefont {Liu}}, \bibinfo {author} {\bibfnamefont {W.}~\bibnamefont {Zhao}},\ and\ \bibinfo {author} {\bibfnamefont {M.}~\bibnamefont {Ezawa}},\ }\bibfield  {title} {\bibinfo {title} {Control and manipulation of a magnetic skyrmionium in nanostructures},\ }\href@noop {} {\bibfield  {journal} {\bibinfo  {journal} {Physical Review B}\ }\textbf {\bibinfo {volume} {94}},\ \bibinfo {pages} {094420} (\bibinfo {year} {2016})}\BibitemShut {NoStop}%
\bibitem [{\citenamefont {Yang}\ \emph {et~al.}(2023)\citenamefont {Yang}, \citenamefont {Zhao}, \citenamefont {Wu}, \citenamefont {Chu}, \citenamefont {Xu}, \citenamefont {Li}, \citenamefont {{\AA}kerman},\ and\ \citenamefont {Zhou}}]{yang2023reversible}%
  \BibitemOpen
  \bibfield  {author} {\bibinfo {author} {\bibfnamefont {S.}~\bibnamefont {Yang}}, \bibinfo {author} {\bibfnamefont {Y.}~\bibnamefont {Zhao}}, \bibinfo {author} {\bibfnamefont {K.}~\bibnamefont {Wu}}, \bibinfo {author} {\bibfnamefont {Z.}~\bibnamefont {Chu}}, \bibinfo {author} {\bibfnamefont {X.}~\bibnamefont {Xu}}, \bibinfo {author} {\bibfnamefont {X.}~\bibnamefont {Li}}, \bibinfo {author} {\bibfnamefont {J.}~\bibnamefont {{\AA}kerman}},\ and\ \bibinfo {author} {\bibfnamefont {Y.}~\bibnamefont {Zhou}},\ }\bibfield  {title} {\bibinfo {title} {Reversible conversion between skyrmions and skyrmioniums},\ }\href@noop {} {\bibfield  {journal} {\bibinfo  {journal} {Nature Communications}\ }\textbf {\bibinfo {volume} {14}},\ \bibinfo {pages} {3406} (\bibinfo {year} {2023})}\BibitemShut {NoStop}%
\bibitem [{\citenamefont {G{\"o}bel}\ \emph {et~al.}(2019)\citenamefont {G{\"o}bel}, \citenamefont {Sch{\"a}ffer}, \citenamefont {Berakdar}, \citenamefont {Mertig},\ and\ \citenamefont {Parkin}}]{gobel2019electrical}%
  \BibitemOpen
  \bibfield  {author} {\bibinfo {author} {\bibfnamefont {B.}~\bibnamefont {G{\"o}bel}}, \bibinfo {author} {\bibfnamefont {A.~F.}\ \bibnamefont {Sch{\"a}ffer}}, \bibinfo {author} {\bibfnamefont {J.}~\bibnamefont {Berakdar}}, \bibinfo {author} {\bibfnamefont {I.}~\bibnamefont {Mertig}},\ and\ \bibinfo {author} {\bibfnamefont {S.~S.}\ \bibnamefont {Parkin}},\ }\bibfield  {title} {\bibinfo {title} {Electrical writing, deleting, reading, and moving of magnetic skyrmioniums in a racetrack device},\ }\href@noop {} {\bibfield  {journal} {\bibinfo  {journal} {Scientific Reports}\ }\textbf {\bibinfo {volume} {9}},\ \bibinfo {pages} {12119} (\bibinfo {year} {2019})}\BibitemShut {NoStop}%
\bibitem [{\citenamefont {Seng}\ \emph {et~al.}(2021)\citenamefont {Seng}, \citenamefont {Sch{\"o}nke}, \citenamefont {Yeste}, \citenamefont {Reeve}, \citenamefont {Kerber}, \citenamefont {Lacour}, \citenamefont {Bello}, \citenamefont {Bergeard}, \citenamefont {Kammerbauer}, \citenamefont {Bhukta} \emph {et~al.}}]{seng2021direct}%
  \BibitemOpen
  \bibfield  {author} {\bibinfo {author} {\bibfnamefont {B.}~\bibnamefont {Seng}}, \bibinfo {author} {\bibfnamefont {D.}~\bibnamefont {Sch{\"o}nke}}, \bibinfo {author} {\bibfnamefont {J.}~\bibnamefont {Yeste}}, \bibinfo {author} {\bibfnamefont {R.~M.}\ \bibnamefont {Reeve}}, \bibinfo {author} {\bibfnamefont {N.}~\bibnamefont {Kerber}}, \bibinfo {author} {\bibfnamefont {D.}~\bibnamefont {Lacour}}, \bibinfo {author} {\bibfnamefont {J.-L.}\ \bibnamefont {Bello}}, \bibinfo {author} {\bibfnamefont {N.}~\bibnamefont {Bergeard}}, \bibinfo {author} {\bibfnamefont {F.}~\bibnamefont {Kammerbauer}}, \bibinfo {author} {\bibfnamefont {M.}~\bibnamefont {Bhukta}}, \emph {et~al.},\ }\bibfield  {title} {\bibinfo {title} {Direct imaging of chiral domain walls and n{\'e}el-type skyrmionium in ferrimagnetic alloys},\ }\href@noop {} {\bibfield  {journal} {\bibinfo  {journal} {Advanced Functional Materials}\ }\textbf {\bibinfo {volume} {31}},\ \bibinfo {pages} {2102307} (\bibinfo {year} {2021})}\BibitemShut {NoStop}%
\bibitem [{\citenamefont {Cartwright}\ and\ \citenamefont {Piro}(1992)}]{cartwright1992dynamics}%
  \BibitemOpen
  \bibfield  {author} {\bibinfo {author} {\bibfnamefont {J.~H.}\ \bibnamefont {Cartwright}}\ and\ \bibinfo {author} {\bibfnamefont {O.}~\bibnamefont {Piro}},\ }\bibfield  {title} {\bibinfo {title} {The dynamics of runge--kutta methods},\ }\href@noop {} {\bibfield  {journal} {\bibinfo  {journal} {International Journal of Bifurcation and Chaos}\ }\textbf {\bibinfo {volume} {2}},\ \bibinfo {pages} {427} (\bibinfo {year} {1992})}\BibitemShut {NoStop}%
\bibitem [{\citenamefont {Chen}\ \emph {et~al.}(2021)\citenamefont {Chen}, \citenamefont {Wan},\ and\ \citenamefont {Zhan}}]{chen2021engineering}%
  \BibitemOpen
  \bibfield  {author} {\bibinfo {author} {\bibfnamefont {J.}~\bibnamefont {Chen}}, \bibinfo {author} {\bibfnamefont {C.}~\bibnamefont {Wan}},\ and\ \bibinfo {author} {\bibfnamefont {Q.}~\bibnamefont {Zhan}},\ }\bibfield  {title} {\bibinfo {title} {Engineering photonic angular momentum with structured light: a review},\ }\href@noop {} {\bibfield  {journal} {\bibinfo  {journal} {Advanced Photonics}\ }\textbf {\bibinfo {volume} {3}},\ \bibinfo {pages} {064001} (\bibinfo {year} {2021})}\BibitemShut {NoStop}%
\end{thebibliography}%

\end{document}